\documentclass[%
 aps,
 pra,
 floatfix,
 amsmath, amssymb,
 reprint,%
]{revtex4-1}

\usepackage[utf8]{inputenc}
\usepackage{hyperref}

\usepackage{graphicx}% Include figure files
\graphicspath{ {./figures/} }
\usepackage{dcolumn}% Align table columns on decimal point
\usepackage{bm}% bold math
\usepackage{braket}

\usepackage{xcolor}

\begin{document}

\preprint{APS/123-QED}

\title[Spatially-selective in situ magnetometry of ultracold atomic clouds
]{Spatially-selective in situ magnetometry of ultracold atomic clouds}

% Spatially-resolved probing of ultracold atomic clouds for magnetometry

%\footnote{Error!}}% Force line breaks with \\
%\thanks{Footnote to title of article.}

\author{Ottó Elíasson}
% \email{ottoel@phys.au.dk}
 \affiliation{ 
Institut for Fysik og Astronomi, Aarhus Universitet, 8000 Aarhus C, Denmark
}%

\author{Robert Heck}
 \affiliation{ 
Institut for Fysik og Astronomi, Aarhus Universitet, 8000 Aarhus C, Denmark
}%

\author{Jens S.\ Laustsen}
 \affiliation{ 
Institut for Fysik og Astronomi, Aarhus Universitet, 8000 Aarhus C, Denmark
}%

\author{Mario Napolitano}
 \affiliation{ 
Institut for Fysik og Astronomi, Aarhus Universitet, 8000 Aarhus C, Denmark
}%

\author{Romain Müller}
 \affiliation{ 
Institut for Fysik og Astronomi, Aarhus Universitet, 8000 Aarhus C, Denmark
}%

\author{Mark G.\ Bason}
 \affiliation{ 
Department of Physics and Astronomy, University of Sussex, Falmer, Brighton BN1 9QH, United Kingdom
}%

\author{Jan J.\ Arlt}
 \affiliation{ 
Institut for Fysik og Astronomi, Aarhus Universitet, 8000 Aarhus C, Denmark
}%

\author{Jacob F.\ Sherson}%
 \email{sherson@phys.au.dk}
\affiliation{ 
Institut for Fysik og Astronomi, Aarhus Universitet, 8000 Aarhus C, Denmark
 %\\This line break forced with \textbackslash\textbackslash
}%

% \homepage{http://www.Second.institution.edu/~Charlie.Author.}

\date{\today}% It is always \today, today,

\begin{abstract}
%1) Physics
%	a) Faraday
%	b) Camera
%We demonstrate novel implementations of high-precision optical magnetometers which allow for spatially selective \emph{and} resolved in situ probing of atomic clouds. This is realised by using shaped dispersive probe beams combined with spatially resolved balanced homodyne detection. Two magnetometer sequences are discussed: first, a vectorial magnetometer
%with single-shot precision of $\delta B_{0} = 50~\mu$G and $\delta B_\perp = 100~\mu$G for magnetic field components along and transverse to the probing direction, respectively, improving a previous realisation by two orders of magnitude. Second, a Larmor magnetometer capable of measuring absolute magnetic fields to a precision of $\delta B = 15~\mu$G. We characterise the dependency of the single-shot precision on the size of the analysed region of interest. By shaping the probe beam, we show the selective probing of individual atomic clouds. 
%%Operating the Larmor magnetometer on an atom cloud array, 
%We can assign a lower bound to the measurement precision of magnetic field gradients of $\frac{dB}{dy}=1.5~\mu$G$/\mu$m.
%Finally, we give an outlook on how dynamic trapping potentials combined with selective probing can be used to realise enhanced quantum simulations in our newly established quantum gas microscope. By defining separate sensing regions in direct proximity to an interaction region, magnetic fields can be measured while the complex quantum dynamics remains unaffected.

%% Alternative version without numbers:
We demonstrate novel implementations of high-precision optical magnetometers which allow for spatially-selective \emph{and} spatially-resolved in situ measurements using cold atomic clouds. These are realised by using shaped dispersive probe beams combined with spatially-resolved balanced homodyne detection. Two magnetometer sequences are discussed: a vectorial magnetometer,
which yields sensitivities two orders of magnitude better compared to a previous realisation and a Larmor magnetometer capable of measuring absolute magnetic fields. 
We characterise the dependence of single-shot precision on the size of the analysed region for the vectorial magnetometer and provide a lower bound for the measurement precision of magnetic field gradients for the Larmor magnetometer. 
%By shaping the probe beam, we show the selective probing of individual atomic clouds. 
%Operating the Larmor magnetometer on an atom cloud array, 
Finally, we give an outlook on how dynamic trapping potentials combined with selective probing can be used to realise enhanced quantum simulations in quantum gas microscopes.% By defining separate sensing regions in direct proximity to an interaction region, magnetic fields can be measured while the complex quantum dynamics remains unaffected.

\end{abstract}

\pacs{Valid PACS appear here}% PACS, the Physics and Astronomy
%Use showkeys class option if keyword
\maketitle

%%%%%%%%%%%%%%%%%%%%%%%%%%%
%%% Introduction
%%%%%%%%%%%%%%%%%%%%%%%%%%%
\section{Introduction} 
Magnetometry has applications in a number of disciplines ranging from medicine to geophysics~\cite{budker_Optical_2013}.
High precision magnetometry also plays a central role in fundamental physics research, for instance in the search for 
a permanent electron electric dipole moment or signs of CPT violation~\cite{safronova_Search_2018}. 

State-of-the-art magnetometers are based on a variety of different methods.
These prominently include superconducting quantum interference devices, solid state Hall sensors, magnetic resonance force microscopes, NV centres and optical magnetometers (OPMs)~\cite{budker_Optical_2013,[{{See supplementary information of} }] [{ {. This gives an excellent summary of the achieved sensitivities in different experimentally realised magnetometers across the different available technologies.}}]muessel_Scalable_2014}.

%{\color{red} The supplementary information of Ref.~\cite{muessel_Scalable_2014} gives an excellent summary of the achieved sensitivities in different experimentally realized magnetometers across the different available technologies.}

OPMs are typically based on the interaction of a laser beam with polarised atomic media. 
In particular vapour-cell-based Spin-Exchange-Relaxation-Free (SERF) optical magnetometers 
offer the highest sensitivity to date~\cite{kominis_subfemtotesla_2003,dang_Ultrahigh_2010}. 
A special subgroup of OPMs are cold atom based magnetometers, which are at the focus of this work~\cite{vengalattore_HighResolution_2007, sewell_Magnetic_2012}. 
These devices have applications in laboratory based research, and their advantage lies in the combination
of high sensitivity and high spatial resolution.
This has been utilised for the detection of magnetic field gradients close to
$100$~nG/$\mu$m~\cite{higbie_Direct_2005,koschorreck_High_2011,muessel_Scalable_2014a}.

%%% schematic %%%
% The compilation in 
% folder: figstouse/setup_schem_v7.pdf
% The 5 cloud array: /Calculations and analysis/Mirek_magentometer/160905_dimMag_benchmark/schematic_fig_5cloud_180809.pdf
% script: paper_analysis_mirekMag_180508.m
\begin{figure}[ht]
\centering
\includegraphics[width=1\columnwidth]{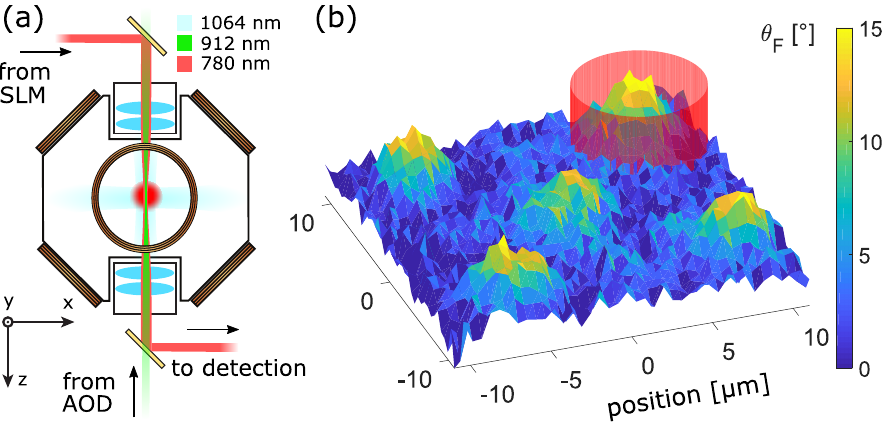}
\caption{(a) Schematic of the optical setup and lasers used in our experiment. A detailed description is in the main text. (b) Our local dispersive probing scheme. The image is a raw dispersive image of the atomic distribution in a system of five microtraps. A red cylinder encircles one of five atomic clouds, as to illustrate how the vectorial dispersive imaging can be made to selectively probe a portion of the atomic system.}
\label{fig:schematic}
\end{figure}
%%%%%%%%%%%%%%%%%%%%%%

In a parallel development, spatial resolution and manipulation of ultra-cold atoms has been
%In future applications it is desirable to allow for spatially-selective measurements in such cold atom based magnetometers. \todo{Any justification needed here?} This spatial resolution can be
realised by employing {\em spatial light modulators} (SLMs) or {\em acousto-optical deflectors} (AODs) {in order to create arbitrarily shaped light fields}. 
{AODs deflect a single laser beam depending on an applied radio frequency (RF) signal. More complex light patterns therefore need to be ``painted'' in a time-averaging approach~\cite{Henderson2009}, or by applying multi-tone RF signals~\cite{endres_Atombyatom_2016}. %not sure that that is the first multi-tone, maybe a better citation could be found
AODs have not only been employed to transport~\cite{kaufman_Entangling_2015} and rearrange~\cite{barredo_atombyatom_2016, endres_Atombyatom_2016} single atoms in microtrap arrays but also in a multiplexed quantum memory module for writing and reading atomic states~\cite{pu_Experimental_2017}.  SLMs modify the phase and/or the amplitude of a laser beam spatially. Their usage in cold atom experiments has steadily been increasing since the first successful atom trapping with an SLM \cite{bergamini_Holographic_2004}, where the traps were generated by a phase modulation. Alternatively, potentials can be created by directly imaging the surface of an SLM onto atoms~\cite{brandt_Spatial_2011,muldoon_Control_2012}. %maybe insert a sentence here introducing DMDs...
% SLMs liquid crystal devices which allow to change the phase [cite] and thus by holography create a potential. There are DMDs which are binary SLMs and can be both used in direct imaging [cite] as well as Fouriedimaging [cite specifically atoms]. 
% Then C.J. Foot and his BEC movement experiment, PRA from 2006
% First DMD trapping is A. Kuhn from 2011 Applied Physics B (proposal!), also NJP 2012 control and manipulation in tweezers (experiment!), and the new one on the moving atoms with DMD's.
Recently, SLMs have been employed in quantum gas experiments to create ultra-low-entropy many-body quantum states \cite{mazurenko_coldatom_2017}. For a more extensive overview of the different implementations the interested reader is referred to~\cite{Holland2018}.}
% Could also cite Fatemi for the SLM of 2007, but there is also an AOD. This is Optics Experss. There are some citations to AOD people, could copy some of those. 

% Reading the Duan experiment, I notice that they don't to any trapping. The experiment is performed in a free falling cloud from a MOT. Aboput 30 µK hot (after polarisation gradient cooling). The experiments are then just performed in the 10's of µs after the MOT is turned off. The "memory cell" are then just the atoms that are in the ensemble, locally inside the read and write beams. So there is no additional trapping, or microtrapping involved as I thought when I read it the first time.

In this paper we describe techniques for spatially-selective and spatially-resolved, single-shot,
in situ magnetometry with cold atoms. 
We discuss the operation of two types of magnetometers: {First, a {\em vectorial magnetometer} which is capable of measuring the longitudinal field component along the probe beam direction, as well as the modulus of the two perpendicular field components in a single experimental realisation. This type of implementation offers good precision. Second, we realise a {\em Larmor magnetometer} which is sensitive to the modulus of the projection of the total field onto the probe axis and provides very high accuracy.} 
Fig.~\ref{fig:schematic}~(a) shows the main features of the experiment. 
The atomic cloud is held in optical microtraps, generated with an AOD, and probed with dispersive light shaped with an SLM. 
The dual objective system in the apparatus allows for manipulation of the atomic cloud in optical tweezers, spatial shaping of the dispersive probe, and spatially-resolved detection of the atomic distribution.

Fig.~\ref{fig:schematic}~(b) depicts an array of five atomic clouds recorded in situ with {dispersive imaging and shows the versatility of the microtrap setup}. The spatial shaping of the dispersive probe beam opens the 
possibility of spatially-selective measurements of parts of the whole atomic system, as indicated by the red cylinder in Fig.~\ref{fig:schematic}~(b), leaving the other regions unaffected. 
%Around one of the clouds is a cylinder in red, indicating minimal size of a probing beam that can be created. 

The dispersive nature of the measurement
makes both magnetometer sequences applicable for in-sequence, in situ measurements of a shot-to-shot fluctuating magnetic field. 
Such fluctuations would otherwise be detrimental for any detailed investigation of quantum spin dynamics such as quantum simulations of spin-chains \cite{simon_Quantum_2011}.
%Such fluctuations would otherwise reduce the quality of a subsequent magnetic field sensitive measurement \todo{any example?}. 
% An atomic system could be split into arbitrarily many spatially distributed \emph{sensor regions} where the magnetic field is sensed, and an \emph{interaction region} where magnetic field sensitive dynamics take place. %This allows for high precision sensing of the magnetic field, and that knowledge can then be applied to compensate for the magnetic field dependent atomic dynamics, that take place in the interaction region.
In the outlook we present our vision for how spatially-selective magnetic field detection in one portion of a system 
can enhance the precision of a magnetic field sensitive measurement in another part, in the context of a new quantum gas microscope experiment.

%\todo{Do we say this is enough, or add the 3 ingredients here? (1) Local dispersive probing (2) Dynamic potential control (3) Quantum Simulator/QGas microscope}.
% In addition we show that the local nature of the scheme paves the way for spatially resolved measurements 
% in the same atomic system. 

% \todo{Add the point that we in fact calibrate the vectorial one with the Larmor one.}

The structure of this paper is as follows. Sec.~\ref{sec:exp} presents the experimental setup,
% the creation of cold thermal clouds of $^{87}$Rb; the generation of microtrap arrays with an AOD; and a balanced homodyne polarimeter based on the Faraday interaction, where the probe light is shaped with a DMD. 
then Secs.~\ref{sec:vec} and \ref{sec:lar} describe the implementation of the two optical magnetometers.
% one based on monitoring the Faraday signal of the atomic cloud, while an external magnetic field is swept through the zero-field. The other is a more typical OPM, where the collective spin state is put into Larmor precession and the Larmor frequency is measured. 
Finally Sec.~\ref{sec:loc} demonstrates the local nature of the dispersive measurement scheme.

% to NV-center based magnetometers \cite{balasubramanian_Nanoscale_2008,maze_Nanoscale_2008} 
% Then we make some overview of the field. Q is how to make that list. We could do it by group, by method, by temperature (some good sorting criterion. Another way is to take it through the sensitivity-volume graph in the \cite{muessel_Scalable_2014a}, supp. mat.

%%%%%%%%%%%%%%%%%%%%%%%%%%%
%%% Experimental setup
%%%%%%%%%%%%%%%%%%%%%%%%%%%

\section{Experimental setup}\label{sec:exp}
The magnetometer sequences discussed here are based on atoms in optical dipole potentials. Our experimental setup was 
previously described in~\cite{bason_Measurementenhanced_2018} and the relevant steps are outlined below.
An experimental cycle is initiated by collecting a cloud of $^{87}$Rb atoms in a $3$D magneto-optical trap.
It is subsequently optically pumped into the 
$\ket{F = 2, m_F =2}$ state and magnetically transported to an intermediate vacuum chamber where
forced microwave evaporation is performed~\cite{bason_Measurementenhanced_2018}. Thereafter the cloud is loaded into a focused single beam optical dipole trap at a wavelength of $1064$~nm, with a $1/e^2$ waist of $45~\mu$m~\cite{Heck2018}. The cloud is then
transported by moving the focus of the beam to a science chamber. Here, another laser beam, 
with a waist of $90~\mu$m, propagating at a 
right angle to the transport beam,
forms a crossed dipole trap (CDT).
Within the CDT we perform further evaporative cooling and obtain 
a thermal atomic cloud of about $2\cdot10^6$ atoms at a temperature of $800$~nK. 

The magnetic field at the position of the atoms is controlled by 
three orthogonal pairs of compensation coils as shown in Fig.~\ref{fig:schematic} (a). 
Fast control of these coils 
is essential for the realisation of the 
magnetometer described below. The effect of $50$~Hz noise is minimised by triggering the
sequence on the power line.

To allow for high-resolution imaging, the science chamber has two 
re-entrant viewports on opposing sides. They are equipped with objectives with an effective numerical aperture NA~$=0.11$.
One of the objectives is used to create optical potentials at high-resolution using
laser light at $912$~nm (shown in Fig.~\ref{fig:schematic} (a)). 
For flexible spatial control this light is sent through a 
two-axis AOD, enabling us to realise time-averaged potentials such as multiple, tightly focused beams. These optical tweezers have a waist of $4.3(1)~\mu$m leading to potential depths in individual tweezers of up to $100~\mu$K.

The dispersive light matter interaction can be decomposed to first order into scalar and vectorial components. 
The typical imaging techniques based on this interaction all have a similar signal-to-noise-ratio~\cite{gajdacz_Nondestructive_2013}. 
Here we make use of the vectorial part that gives access to the magnetisation of the atomic cloud.
This imaging technique is traditionally called Faraday imaging. 
It has commonly been measured with photodetectors, e.g. for magnetometry or spin-squeezing experiments~\cite{isayama_Observation_1999,sewell_Magnetic_2012, palacios_Multisecond_2018}, however these experiments lack the ability to obtain spatial information of the atomic clouds. Recently, spatially-resolved Faraday imaging has enabled studies of density effects in the light-matter interaction~\cite{kaminski_Insitu_2012}, in situ dynamics of thermal atomic clouds~\cite{gajdacz_Nondestructive_2013}, sub shot-noise atom number stabilised ultracold clouds~\cite{gajdacz_Preparation_2016} and the BEC phase transition itself~\cite{bason_Measurementenhanced_2018}. 

\begin{figure}[tbp]
\centering
\includegraphics[width=1\columnwidth]{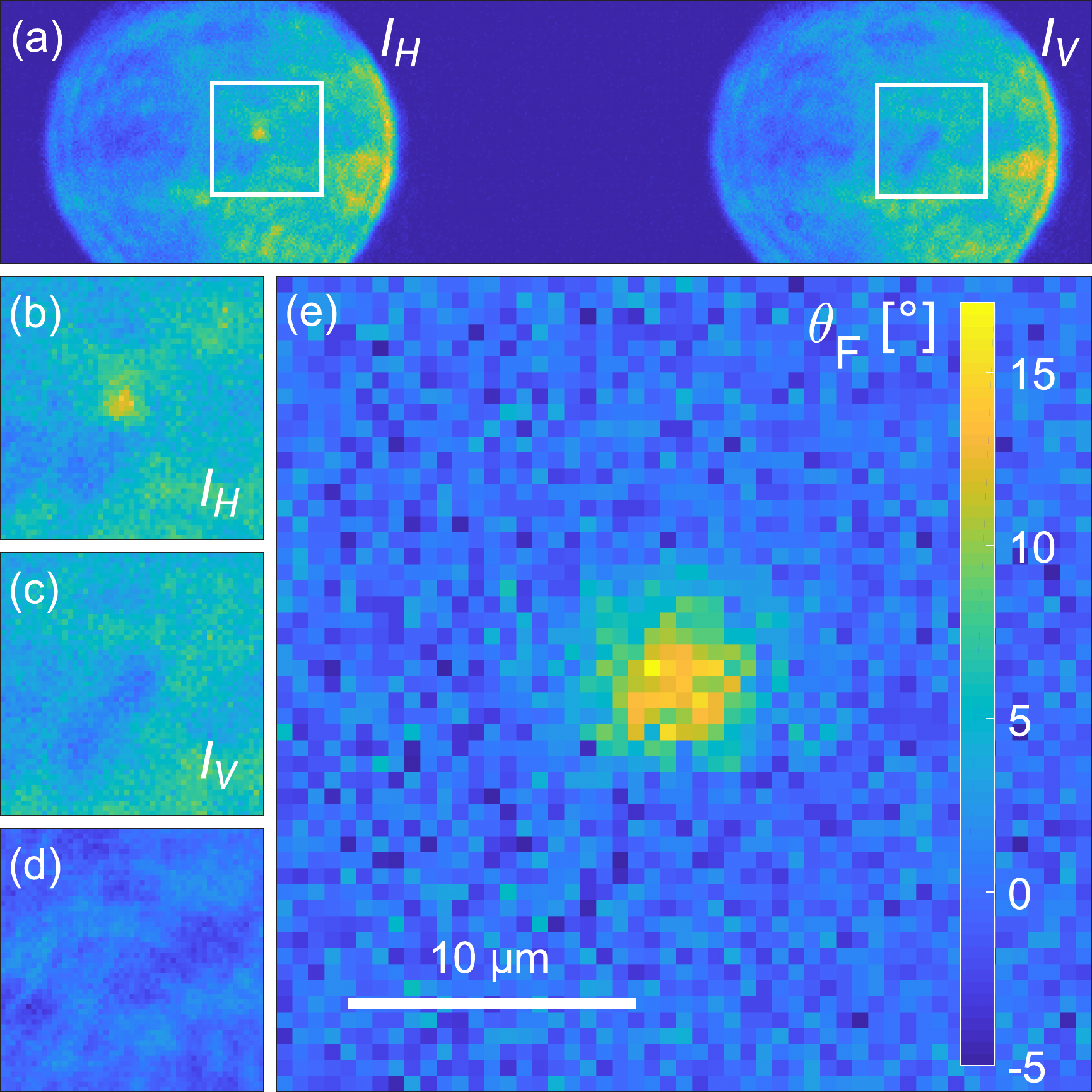}
\caption{Dual Port Faraday Imaging. Panel (a) displays the raw polarisation image obtained on the camera. The two white frames enclose the area shown in panels (b) and (c) which contain the raw polarisation fields $I_{\rm H}$ and $I_{\rm V}$ respectively. The colourscale in panels (a)-(c) represent the light intensity. Panel (d) is an averaged Faraday image without atoms and (e) contains the final Faraday image. The colourscale of panel (d) is the same as that of (e).}
%\caption{Dual Port Faraday Imaging. Panels (a) and (b) contain the raw polarisation fields $I_{\rm H}$ and $I_{\rm V}$ respectively. \todo{What is really the colourscale here? Maybe better to show the full beam here, to see that it goes to zero. Say this was a flat top beam with so and so modulation on top.} Panel (c) is an averaged background image and (d) contains the final Faraday image.}
\label{fig:homodyne}
\end{figure}

Here, we use the Faraday interaction to detect the atoms non-destructively. %apply the Faraday detection light with spatial resolution.
Briefly, a linearly polarised, off-resonant probe beam is sent through the atomic sample (shown in Fig.~\ref{fig:schematic}~(b)). 
{Given that all atoms are prepared in the same internal state, t}he plane of polarisation of the probe is rotated by the Faraday angle $\theta_F$
according to $\theta_F(x,y)\sim\braket{f_z}\int \rho(x,y,z)\,dz$, where $\braket{f_z}$ is the {average} projection of the total angular momentum state {per atom} along the probing direction and {$\rho(x,y,z)$} corresponds to the {atom} density \cite{gajdacz_Nondestructive_2013,bradley_BoseEinstein_1997}. {In the limit of large atom numbers $\braket{f_z}$ can be treated as a classical quantity.
%A novel feature of our setup is the ability to shape the Faraday probe beam. 

To achieve spatially-selective imaging, Faraday light is shaped by imaging the plane of a SLM (in our experiment a {\em Digital Micromirror Device} (DMD)) onto the atoms.}
%, thereby gaining arbitrary control over the spatial structure of the probe light.
After the chamber 
the Faraday light passes through a balanced homodyne polarimeter~\cite{kaminski_Insitu_2012}.
This polarimeter consists of a Wollaston prism which separates light into
the horizontal and vertical polarisation components.
These components are then imaged through the same imaging system 
onto {separate areas of} an {\em electron multiplying charge-coupled device} (EMCCD) camera ({\sc Andor} iXon Ultra 897). {The resulting images are denoted as $I_{\rm H}$ and $I_{\rm V}$, corresponding to the horizontal and vertical polarimeter ports, respectively.}
A Faraday image {describes} the spatially dependent angle $\theta_F$ and is generated %from these images 
by combining $I_{\rm H}$ and $I_{\rm V}$ according to~\cite{kaminski_Insitu_2012}
\begin{align}
\theta_F = \frac{1}{2}\arcsin{ \left( \frac{I_{\rm H}-I_{\rm V}}{I_{\rm H}+I_{\rm V}} \right) },\label{eq:fara}
\end{align}
for each camera pixel{. For optimal performance, careful polarisation balancing of $I_{\rm H}$ and $I_{\rm V}$ is crucial.} 
%This is achieved by a correct relative positioning of the two polarization fields when applying Eq.~(\ref{eq:fara}).} 
In Fig.~\ref{fig:homodyne}~(a) a raw image of the two polarisation fields is shown. Here the Faraday light is shaped as a flat top beam, but the light intensity varies somewhat due to imperfections in the imaging system. {These variations will lead to increased technical noise in the Faraday image if $I_{\rm H}$ and $I_{\rm V}$ are not correctly centred with respect to each other. Using this noise as a measure allows one to find the right relative positioning of the images through a minimisation algorithm.}
% - careful balancing of the two ports is necessary and done with a L/2 waveplate; small remaining imbalances can be corrected in the post processing of the images
% - correct centring of the images is crucial. we use light patterns generated with the DMD in images without atoms. as we use the same optics for imaging and there are no significant differences in the optical path a simple minimization algorithm can be used to find the right pixel shifts for the correct cancellation
Figs.~\ref{fig:homodyne}~(b) and (c) 
are enlarged regions of the raw polarisation fields $I_{\rm V}$ and $I_{\rm H}$ around the position of the atoms.
%, and (c) shows the result of applying eq.~(\ref{eq:fara}) to the two fields. 
To account for {remaining}
image artifacts originating in the optical setup, an averaged Faraday image without atoms (d) is subtracted.
The final {Faraday} image in Fig.~\ref{fig:homodyne}~(e) shows an atomic cloud trapped in a single optical tweezer, with a 
peak rotation angle of around $\theta_F\simeq 15^\circ$. 
%The signal-to-noise ratio of a given signal is intimately related to destructivity and the optical depth (OD) of the sample; typically here the OD~$\simeq2.5$.
The main advantage of minimally-destructive dispersive imaging methods is the option to take multiple 
images in a single realisation. This enables us to investigate dynamics in the atomic system  \cite{gajdacz_Nondestructive_2013,bason_Measurementenhanced_2018}.

%%% homodyne images %%%
% script: Calculations and analysis/FaradayImg_compilation/FaradayImg_compilation.m
%%%%%%%%%%%%%%%%%%%%%%

%%% Mirek benchmark %%%
% (a): script:/Calculations and analysis/160602_resMag_benchmark_v3/prelimMultiFaradaySuperplot_wError_160602.m
% (b): script:/Calculations and analysis/160615_dimMag_overnight/tempTraceForDelA_170319.m
% compilation: figstouse/vecMag_compilation_v3.pdf
\begin{figure*}[htp]
\centering
\includegraphics[width=2\columnwidth]{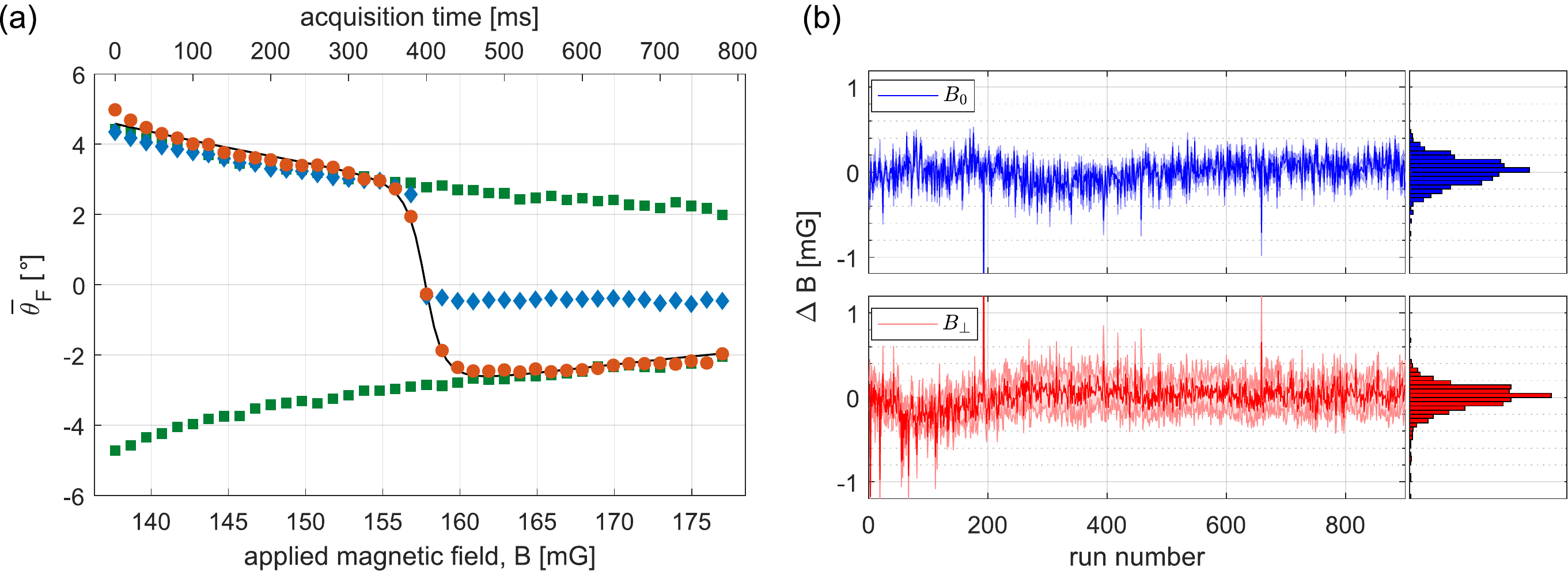}
\caption{Vector magnetometry. (a) A typical trace of a single magnetometer sequence as a function {of the applied magnetic field} is shown in red ({\large$\bullet$}). {The data are fit using eq. (\ref{eq:mirek_trace}), and the result is shown as the black solid line.} The blue trace ($\blacklozenge$) occurs when the background fields are compensated below the threshold for Majorana losses. The green ({\tiny$\blacksquare$}) curves show the intrinsic signal loss during probing; they are produced by keeping the bias field at a fixed value. {The top axis shows the acquisition time of the magnetic field sweep. The absolute calibration of the applied field was obtained using the Larmor magnetometer presented in Sec.~\ref{sec:lar}}. (b) Long term magnetic field drift. The magnetometer was operated continuously for about 900 experimental runs and the magnetic field values extracted for each run, $B_{0}$ (blue) and $B_{\perp}$ (red). {The shaded areas are $1\sigma$ errors of the fits. The histograms are rendered from the fit results.}}\label{fig:mirek_benchmark}
\end{figure*}
%%%%%%%%%%%%%%%%%%%%%
 
%%%% Mirek benchmark %%%
%\begin{figure}[htbp]
%\centering
%\includegraphics[width=1\columnwidth]{figstouse/mirekMag_benchmark.pdf}
%\caption{\todo{Separate frames, or at least have 2 y axes. Here I must somehow change the horizontal axis, this is not honest. It has to be time, as long as I have the decaying traces. Maybe I could then put an auxillary axis as the real field, the coil current, or the applied field depending on the point I want to make.
%This could be the cloud at different points in the the sweep e.g. Put different markers for 
%the colour blind on the plot}}\label{fig:mirek_benchmark}
%\end{figure}
%%%%%%%%%%%%%%%%%%%%%%

%%%%%%%%%%%%%%%%%%%%%%%%%%%
%%% Magnetometry: vectorial
%%%%%%%%%%%%%%%%%%%%%%%%%%%
\section{Vectorial magnetometry}\label{sec:vec}
%%%% Mirek trace %%%
%\begin{figure}[htbp]
%\centering
%\includegraphics[width=1\columnwidth]{figstouse/mirekMag_trace.png}
%\caption{The figure could be a trace, and then somehow using the long time scan (900 shots time trace of the Mirek). The spikes there should be real, but one would in principle need to look at the fits to make sure. Good for giving the noise band. Move the inset to the white part.}\label{fig:mirek_trace}
%\end{figure}
%%%%%%%%%%%%%%%%%%%%%%%  

% %%% Cut out this paragraph in interation _v3 from Jan.
% The orientation of the magnetisation of a spin polarised atomic sample follows the 
% instantaneous direction of the local magnetic field, in the adiabatic limit. By taking Faraday images of the atomic cloud and
% simultaneously applying an external magnetic field, one can monitor when the Faraday signal vanishes.
% The applied magnetic field is then equal to the background magnetic field, and has opposite direction.
% By sweeping the external magnetic field, and measuring continuously the Faraday signal, one can also 
% gain knowledge of the magnetic field strength in the transversal directions, so this
% method realises a vectorial magnetometer.
% %%%

In a first set of experiments, a vectorial magnetometer was implemented by taking Faraday images of an 
atomic cloud while the magnetic field is changed in amplitude and direction. {As long as the field sweep is slow compared to the instantaneous Larmor frequency, the orientation of the collective spin of the atomic cloud follows the magnetic field direction. Hence, a magnetic field parallel to the Faraday beam will lead to {$\braket{f_z}>0$} and positive $\theta_F$}. Likewise an antiparallel field component will {result in} $\braket{f_z}<0$ and negative $\theta_F$. When the parallel field component vanishes, $\theta_F$ is zero. % and the width of the zero crossing reveals information on the transverse field components. 

The magnetometer sequence is implemented as follows: 
{After evaporative cooling of the atomic cloud in the CDT, the collective atomic spin is prepared either parallel or antiparallel to the probing direction by raising an offset field within $200$~ms in the respective direction. Subsequently, a} train of $40$ Faraday pulses is triggered. Each pulse is $2~\mu$s in duration and has 
an intensity of $150$~pW$/\mu$m$^2$. The light is $1.13$~GHz blue detuned with respect to
the $F = 2 \rightarrow F'=3$ transition of the $^{87}$Rb D2 line. 
% The pixel to pixel fluctuations in the Faraday rotation of a background image amounts to $0.4^\circ$.
% \todo{Maybe this point is unnecessary.}
% See p. 36 of the BlackBook notebook starting in March 2018. I have 20 bkg images, take the std of each, and then the mean of those std's. Get dtheta = 0.34 deg, which is a measure of the bkg noise level.
The images are acquired at a rate of $f_{\rm aq} = 50$~Hz, such that the total measurement time
is $780$~ms. This rate is chosen
to minimise the shot-to-shot jitter due to the $50$~Hz line frequency. Simultaneously the magnetic field component along the Faraday beam is swept through $\Delta B = 40$~mG at a sweep speed 
of $50~$mG/s. Since the field sweep must be slow compared to 
the instantaneous Larmor frequency, the maximal speed is set by the residual {transverse magnetic field, $B_\perp$}. 
%For each Faraday image, the mean rotation
%angle within a given region of interest (ROI) is obtained as a function of the acquisition time.
By averaging over a given region of interest (ROI) in the Faraday images, we can obtain a signal evolving as a function of acquisition time, reproducing the dynamics of the magnetic field.

A typical trace is shown in red in Fig.~\ref{fig:mirek_benchmark}~(a), where
the mean rotation angle, $\bar{\theta}_F$,  in a ROI of $20\times40~\mu$m$^2$ was used. {In our case,} the external field is swept at a constant rate leading to a linear relation between the acquisition time (bottom axis) and
magnitude of the applied field (top axis).
 {Initially, the signal falls off slowly due to losses caused by the probe light itself. When the field strength reaches a critical value, the transverse components begin to dominate the total magnetic field vector and the signal drops significantly since the collective atomic spin turns with the field}. The signal changes sign %about midway through the trace, 
% The annoying thing is that the dimple signal is really bad, and the 
% precision we get is much worse thatn from this one. 
% At least when looking at this particular trace, we seem to get the bes result, when taking the
% full fram of 100 by 100 px. Then we reduce it to only the peak, around 30 x 30 px the
% fit precision doubles, and it is a bit awkward to quote a different value than plotted.
% \todo{Check this and decide what is the best thing to do here. It is maybe better to use the peak rotation angle,
% or somehow a value from the fitted curves, but this seems to have a negative effect on the fit precision}. 
and the zero crossing corresponds to the instance where the field component along the probe direction is zero. At this point the background magnetic field and the applied field, $B_{0}$, are equal and opposite, cancelling each other. %This field, $B_{0}$, is equal and opposite to the field that must be applied to null the total magnetic field at the position of the atoms.
The width of the zero crossing is related to the magnitude of {$B_\perp$}. %As the field is swept further the ensemble polarisation now points in the opposite direction. 

To extract values for the magnetic field components we fit {the measured $\bar{\theta}_F$ as a function of the applied field $B$ with} the expression~\cite{gajdacz_Nondestructive_2013}
\begin{align}
	\bar{\theta}_F(B) = -Ae^{-k{B}}\frac{(B-B_{0})}{\sqrt{(B-B_{0})^2+B_\perp^2}}.\label{eq:mirek_trace}
\end{align}
{This is based on the assumption that the rotation angle adiabatically follows the magnetic field vector. The loss of signal due to the probe itself is modelled by an exponential decay with decay constant $k$ assuming a constant sweep rate of $B$~\footnote{{Note, that one is not necessarily limited to constant sweep rates. Reducing for example the sweep rate around the zero crossing of the Faraday signal could increase the precision of the magnetometer further.}}. The remaining expression corresponds to the projection of the total magnetic field vector, $\mathbf{B} = (B-B_0,B_\perp)$, onto the probe direction. The parallel and perpendicular magnetic field values $B_{0}$ and $B_\perp$ as well as the amplitude $A$ are free fit parameters.
Based on this procedure it is clear that the accuracy of the measured field} values relies on the calibration of the magnetic coils used for the sweep of the applied field.

For the trace shown in Fig.~\ref{fig:mirek_benchmark}, the fit yields values of
$B_0 = 157.87$~mG and $B_\perp = 1.3$~mG, with single-shot $1\sigma$ {precision of
$\delta B_{0} = 0.05$~mG and $\delta B_\perp = 0.1$~mG}.
An implementation of this {particular kind of a cold atom magnetometer, was demonstrated in~\cite{gajdacz_Nondestructive_2013}, inspired by older spectroscopic scanning-field techniques~\cite{dupont-roc_Detection_1969}.  However the measurement in~\cite{gajdacz_Nondestructive_2013} was not sensitive to the polarisation direction of the spin
ensemble.}
%, as only one port of the polarimeter is measured.
Our realisation represents an improvement of the single-shot precision
by two orders of magnitude.

Even though the Faraday probe is more than a GHz off-resonant, it induces signal loss via
heating and transfer to different magnetic states. To characterise this loss we have
performed a Faraday measurement while keeping the value of the magnetic field fixed. These traces are shown in green in
Fig.~\ref{fig:mirek_benchmark}~(a) and correspond to the decay caused by the imaging procedure itself. {The different signs of the traces represent atomic clouds initially prepared with collective spins pointing in parallel (positive) and antiparallel (negative) directions, with respect to the probe beam.} 

When the transverse background fields are well compensated
we see a dramatically different behaviour (Fig.~\ref{fig:mirek_benchmark}~(a), blue trace).
The population of the initial spin state is lost due to Majorana spin flips, 
since the total field becomes close to zero.
A loss in the spin signal is expected when the relative rate of change in the 
magnetic field $|\frac{1}{B}\frac{dB}{dt}|$ becomes higher than the Larmor frequency. 
Hence, for the sweep speeds presented here, the fields must be compensated to within $0.2$~mG. 
This corresponds to the level of the passive compensation of
the background fields in our experiment. 

The vectorial magnetometer was used to measure the background field drifts 
over a period of $8$~hours, as shown in Fig.~\ref{fig:mirek_benchmark}~(b). 
{We compare the magnetic field trace to its mean value using $\Delta B=B(t)-\bar{B}$. The $1\sigma$ deviation
of the histograms of both the parallel and transversal field components amounts to about 
$0.2$~mG, which is a typical performance in an unshielded environment. Most notably, the single-shot precision is better than typical shot-to-shot fluctuations}, which renders the method eligible for improving
measurements of atomic properties that depend crucially on the background magnetic field. In such cases one must measure the background field during the experiment and either actively compensate for its effect \cite{smith_Threeaxis_2011} or post select from the data according to the value measured \cite{bason_Measurementenhanced_2018,krinner_situ_2018}.

% I simply take the list of vales, and find the std of the sample. I try to find the fluctuations from shot to shot (some Allan 
% variance kind of thing). I find a vector of a difference of the entry n and n+1, take the abs, and then mean of all the differences, that should be (at least up to a factor of 2) a measure of the shot to shot fluctuations. This gives 0.23, and just 1 std of the whole sample is 0.24 mG, so I see not much difference. Probably I should divide the first number by 2, as to take half of the difference (which is also what is done in a std. deviation, so then it is even lower, as is to be expected if there are any bigger drifts.

In the context of our experiment, this method {mainly} serves as a probe upon which subsequent measurements or dynamics can be conditioned. {For completeness, we also quote the sensitivities, corresponding to the single-shot magnetic field precision normalized to the total running time of one measurement, as is customary for optical magnetometers} \footnote{{In the field of ultracold quantum gases, magnetic fields are typically given in units of Gauss, and use this convention here, as the paper is primarily intended for that audience. However, this choice of units is not the same in the field of optical magnetometry, where the central figure of merit, the sensitivity, is typically given in SI units (of T~Hz$^{-\frac{1}{2}}$). For this reason we state magnetic fields in units of G, but sensitivities in units of T~Hz$^{-\frac{1}{2}}$.}}. At the full experimental cycle of $30$~s, this yields sensitivities of 
$\delta B_{0}\sqrt{T} = 27$~nT~Hz$^{-\frac{1}{2}}$ and
$\delta B_{\perp}\sqrt{T} = 55$~nT~Hz$^{-\frac{1}{2}}$
% This is based on the 60 nT/sqrt(Hz) stated in the paper and the statement from Mick that 
% the lattice cycle times are around 90 s long.

Thus, 
%this magnetometer is fairly simple to realise experimentally. 
the fields $B_0$ and $B_\perp$ are obtained with a precision that is much lower than shot-to-shot fluctuations in a single realisation of the experiment, despite the simplicity of the measurement scheme. However, its accuracy relies on the quality of the external field calibration.

%Conclusions/improvements on this magnetometer (should not be too long (mini conclusion), move some to final outlook instead if necessary). Compare to others.
%\begin{itemize}
%  \setlength{\itemsep}{0pt}
%  \setlength{\parskip}{0pt}
%\item It is easy to set up this kind of magnetometer. Can be done with a smaller camera, or a photo detector, say.
%\item The precision is below the run to run fluctuations so it may be easily applied to null fields, just as is common to do now with $\mu$-Waves.
%\item Show somehow this is a different technique from the others? Main point is it's vectorial nature. One run, we get both
%		$B_z$ and $B_\perp$.
%\item Sample the transition more precisely, so distribute more measurement points on the transition itself. That
%		should improve the fit precision around that area.
%\item It is only accurate, not precise.
%\end{itemize}
%
%\todo{(Check if we did Mirek in tweezers, or if Mirek was only in the CDT). 
%Check what precision we reach in the tweezer version was. The tweezer presicion is quite shit.}

%%%%%%%%%%%%%%%%%%%%%%%%%%%
%%% Magnetometry: Larmor
%%%%%%%%%%%%%%%%%%%%%%%%%%%

\section{Larmor magnetometry}\label{sec:lar}
In a second set of experiments, a magnetometer based on Larmor precession was implemented.
Generally, a spin polarised atomic ensemble precesses in a magnetic field 
if the polarisation is tilted with respect to the field. {Since the Larmor frequency is $\omega_L=\gamma B$ for small fields,
where $\gamma$ is the gyromagnetic ratio, the magnetic field strength $B$ can be measured accurately by measuring $\omega_L$. 
In our case using the $\ket{F=2,m_F=2}$ state of $^{87}$Rb, the Larmor frequency is given by $\omega_L/B = 2\pi\times1.4~$MHz/G~\cite{steck_Rubidium_2010}.}

Larmor magnetometry has commonly been applied using cold and ultracold atoms. 
Moreover fast phase contrast imaging has been applied to measure the phase acquired by a precessing
BEC in a synthetic magnetic field~\cite{vengalattore_HighResolution_2007}, yielding single-shot 
precision of $9$~nG. High-resolution gradient magnetometry was performed at a 
precision of about $100$~nG$/\mu$m~\cite{higbie_Direct_2005,koschorreck_High_2011,muessel_Scalable_2014}. 
The method has also been used to measure field curvatures of a quadrupolar field in 
multiple experimental realisations on the mm scale \cite{fatemi_Spatially_2010}.
Here we apply an established method of magnetometry and put it in the context 
of spatially-resolved probing that could enable the reduction of detrimental effects 
induced by environmental fluctuations on magnetic field sensitive atomic dynamics.
%Here we perform Larmor magnetometry with a spatially shaped Faraday probe and spatial detection.
% I wrote here something by some guy called Lett. I simply don't remember any papers. Found a researcher called P.D Lett at NIST by some googleing, but not directly doing magnetometry.

%%% Larmor Oscillation %%%
% script: /Calculations and analysis/160817_19_dimMag_larmor_Ivert_scan_v2/19_dimMag_larmor_Ivert_scan_v2/paper_compilation_fdfscan_180425.m
\begin{figure}[htp]
\centering
\includegraphics[width=1\columnwidth]{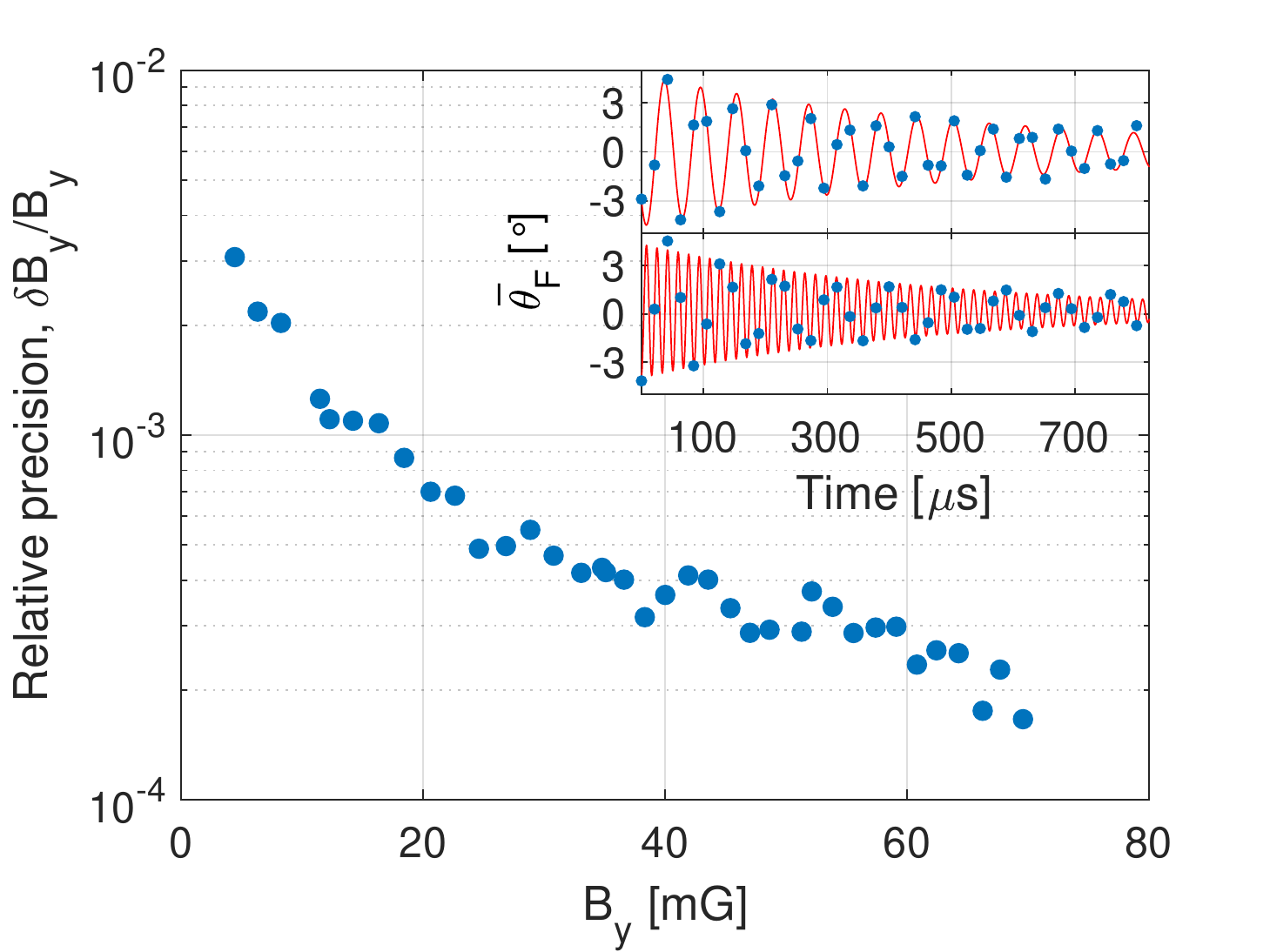}
\caption{Larmor magnetometry. The Larmor magnetometer was tested for a range of magnetic field strengths. The relative precision of the measurement increases with increasing field. The inset shows experimental data for the Larmor magnetometer {as blue dots, and fits as the red solid lines}. In the top panel the measured field was 12.253(13)~mG and in the bottom panel 41.894(17)~mG. {Note that the field of the latter measurement corresponds to Larmor frequencies above $f_{\rm aq}$ which was taken into account in the fit procedure.}}\label{fig:larmor}
\end{figure}
%%%%%%%%%%%%%%%%%%%%%%

%%% Local detection %%%
% (a): script:/Calculations and analysis/160602_resMag_benchmark_v3/paper_analysis_mirekMag_Nslice_Illustration_180911.m
% (b): script: /Calculations and analysis/160602_resMag_benchmark_v3/paper_analysis_mirekMag_Nslice_180430.m
% Note that there is also a _v2. I believe though this is the one I use to produce the plot, there is an addition in the _v2 which was not used in the end.
% compilation: figstouse/localDetProbe_compilation_p1_v2.pdf
\begin{figure*}[htbp]
\centering
\includegraphics[width=1\textwidth]{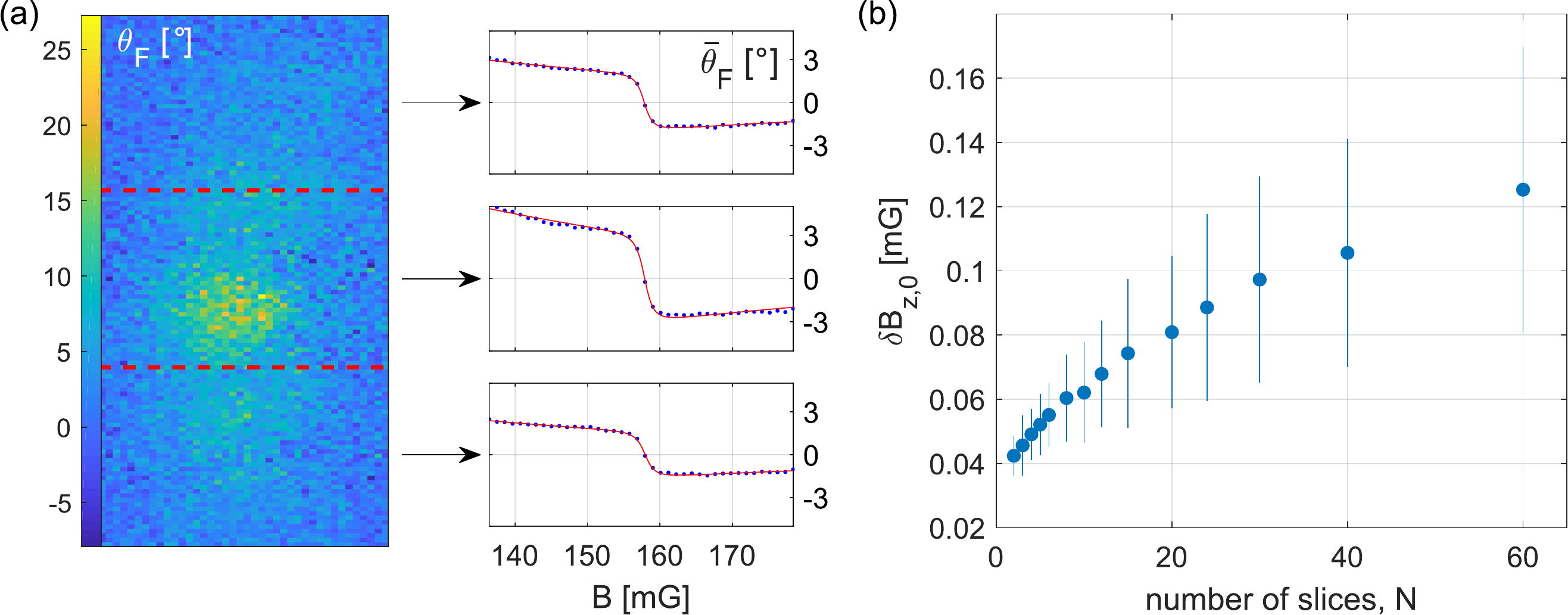}
\caption{Spatially-resolved detection. (a) Individual images of a single run of the vectorial magnetometer are analyzed in multiple slices (here in $N = 3$ slices) along the long extension of the cloud. The $3$ graphs display the mean Faraday angle and the fitted magnetometer trace, and the image is the first frame in the series. (b) The procedure was applied for a varying number of slices $N$ and the average precision as a function of the number of slices $N$ is shown. The error bar represents $1\sigma$ of the $N$ fitted precisions.}\label{fig:detection_local}
\end{figure*}
%%%%%%%%%%%%%%%%%%%%%%

The experimental sequence starts by preparing a cold spin polarised thermal cloud of  $\ket{2,2}$ atoms in the CDT. Thereafter, a 
single additional tweezer is ramped to a potential depth of about 
$2~\mu$K, holding about $1000$ atoms. By doing so, we locally enhance the optical density of the cloud, and thereby allow for an improvement
of the Faraday signal.  {Since these experiments measure an applied magnetic field component along the vertical direction ($y$-axis),
$B_\perp = \sqrt{B_x^2+B_y^2}$ is initially} % maybe if we think at all necessary move the definition of B_\perp to a place somewhere earlier. In any case I think it has to be clear to the reader either explicitly or implicitly that B_\perp contains also the vertical direction
compensated to below $0.2~$mG and the longitudinal field is set to $B_{0}=5$~mG along the probing direction
in order to maintain the magnetisation of the cloud.
To initiate Larmor precession, a magnetic field is abruptly turned on in the vertical direction. 
The initial ramp speed of $5~$mG/$\mu$s is large enough such that the spin polarisation does not
follow, and the collective spin state starts to precess around the $y$-axis.
Thereafter the magnetic field along the $z$-direction is nulled. Precisely $20$~ms later -- one cycle of the $50$~Hz power line -- the field along the $y$-axis is turned on and 
a Faraday pulse train is started. 
By operating the EMCCD in {\em fast kinetic mode} and selecting a ROI of only $25$ pixels
height around the Faraday signal of the atoms, 
we reach an acquisition frequency of $f_{\rm aq} = 44.31(2)$~kHz,
resulting in a total probing time of $T_{\rm aq} = 925~\mu$s. 
The acquisition procedure is similar to the one presented in~\cite{higbie_Direct_2005}.
Typical measurement results are shown in the two insets of
Fig.~\ref{fig:larmor}, where the mean rotation angle in a ROI of $5\times5~\mu m^2$ ($10\times10$ camera pixels) was used to extract the measured quantities. By fitting an exponentially damped 
sinusoidal oscillation to the trace we extract the Larmor frequency. 
Thus, we can determine the magnetic field to a $1\sigma$ precision of
$\delta B_{y} = 15 \ \mu$G in a single realisation of the experiment. Given the full experimental cycle time of $T = 30~$s
this corresponds to a 
sensitivity of $\delta B_{y}\sqrt{T} = 8$~nT~Hz$^{-\frac{1}{2}}$.
% \todo{Should all be in Tesla?}, 
% where the volume of the probed sample is 6000~$\mu$m$^3$.
% Vol is waist * waist * pi * diameter of cloud. Maybe revisit this a bit.

To test the dynamic range of the magnetometer we performed the experiment
for a range of magnetic field strengths from $3$ mG to $70$ mG. The absolute precision is roughly constant over the range of measured field strengths and hence
the relative precision increases, as shown in Fig.~\ref{fig:larmor}.

This method provides an accurate way of measuring the magnetic field at the atomic cloud position. 
It can be used to measure higher magnetic fields than with the vectorial magnetometer covered above in Sec.~\ref{sec:vec}. As a potential, improvement the Larmor precession could be initiated via optical pumping instead of a sudden field change. 
In that way the initial phase of the oscillation can be fixed. Moreover, this method would also allow for a spatially-selective way of initiating the precession. This would be necessary for a magnetometer sequence which uses spatially-selective probing. In the following section we extend the magnetometer sequences and demonstrate the capabilities to perform such measurements.

%%% Points for conclusions
%\todo{I am not sure how to conclude this section, what is a good point to stress. The further improvements, or the usage? Also because it is a DC magnetometer, so using it to map AC background fields seems quite far fetched.}
%
%This type of magnetometer is especially useful for in situ measurements ...
%
%Conclusions/improvements on this magnetometer (should not be too long (mini conclusion), move some to final outlook instead if necessary). Compare to others.
%\begin{itemize}
%  \setlength{\itemsep}{0pt}
%  \setlength{\parskip}{0pt}
%\item Get it working via optical pumping instead of the field kick.
%\item Really check the detection limits in terms of frequency. 
%		I believe we can actually measure at least more than just the $70$~mG.
%\item This is DC field detection.
%\item Point is that one needs to be able to measure the maximal noise component that is possibly present in the 
%		system, i.e. in the same bandwidth as the dynamics one intends to probe in the actual experiment. 
%		Magnetometers benchmarking quantum dynamics.
%\item For it to be really useful beyond the passively stabilised one could combine it with active stabilisation.
%\item Or use it to measure the B-field in situ while other interesting B-field dependent dynamics are happening, 
%		in the spirit of \cite{krinner_situ_2018}.
%\end{itemize}
%
%%%

%%%%%%%%%%%%%%%%%%%%%%
%%%%%%%%%%%%%%%%%%%%%%%%%%%
%%% Probe locality
%%%%%%%%%%%%%%%%%%%%%%%%%%%

\section{Spatial probing and detection schemes}\label{sec:loc}
Our experimental setup offers possibilities of both spatially-resolved detection and spatially-selective probing. 
The spatial resolution of the detection is due to the fact that the Faraday signal is recorded on a camera, where conventionally it is only measured with a 
photo detector. 
% Imaging Faraday signals has been employed before to acquire in situ 
% cloud distributions \cite{bason_Measurementenhanced_2018,kaminski_Insitu_2012, gajdacz_Nondestructive_2013, gajdacz_Preparation_2016}. 
The spatial selectivity in probing is enabled by shaping of the Faraday probe light with a DMD. 

%%% Local probe %%%
% (a) script: /Calculations and analysis/160823_probeLocality/probeScan_analysis_160823.m
% (b) script: /Calculations and analysis/160817_23_dimMag_larmor_multi/paper_compilation_script_180420.m
\begin{figure*}[htbp]
\centering
\includegraphics[width=1\textwidth]{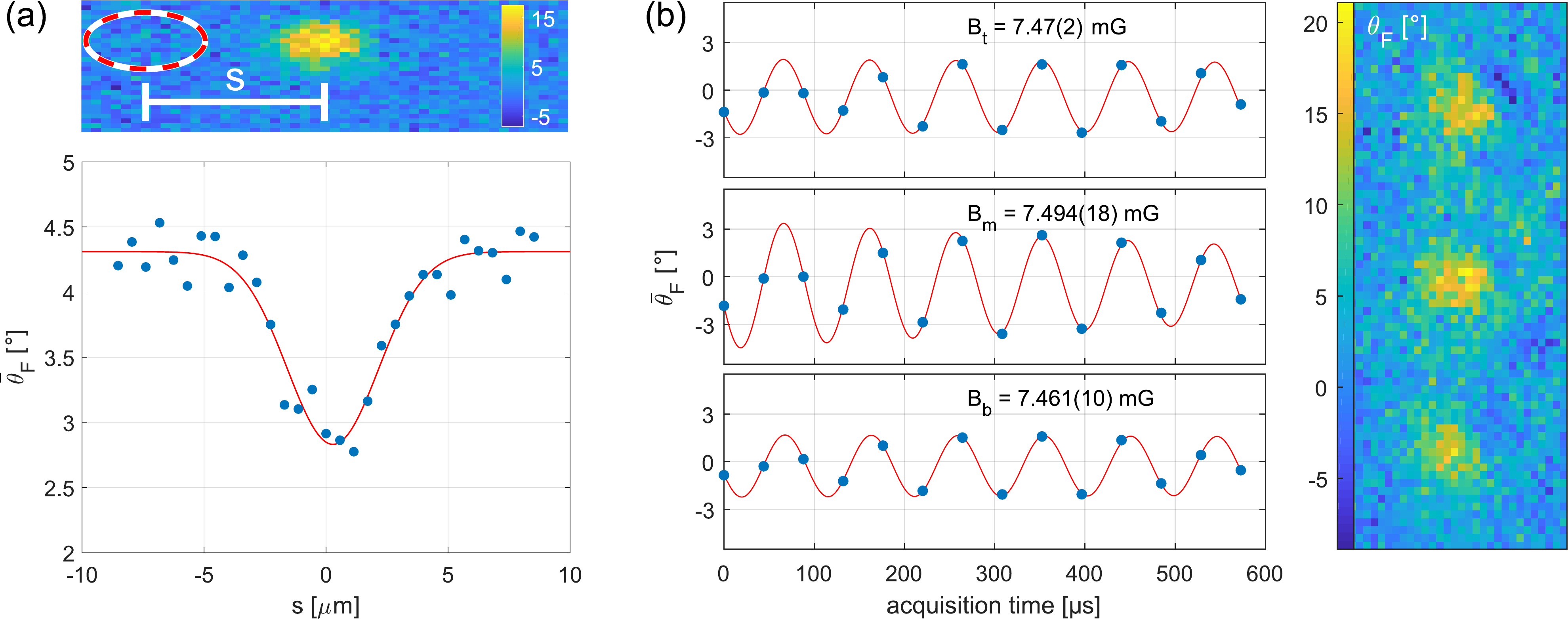}
\caption{Spatially-selective probing. (a) Locality of the probe light. %As a small Faraday probe is scanned across a small atomic cloud contained in a single tweezer, the atomic signal is measured in a second subsequent image.
The atoms are only affected by the DMD shaped probe beam if the probe light overlaps the atomic cloud. (b) Larmor magnetometry in an array. The Larmor magnetometer was applied to atoms in a $3\times1$ array of tweezers, shown in the right panel. The obtained magnetic field values are printed in Larmor precession traces to the left.}\label{fig:probe_local}
\end{figure*}
%%%%%%%%%%%%%%%%%%%%%

Generally, the best signal-to-noise-ratio is obtained when the total Faraday signal from the whole cloud
is recorded. This can be realised by summing the measured Faraday rotation 
pixel by pixel, which is essentially the same as using a photo detector.
Such a situation will however inhibit 
spatially-resolved detection of the magnetic fields in different microtraps, 
as the spatial information in the image is not used. 

To investigate the single-shot precision between the limiting cases of using the full spatial resolution or summing all the signal, we use the measurements taken with the vectorial magnetometer, shown in Fig.~\ref{fig:mirek_benchmark}~(a). The dataset consists of $40$ Faraday images with a size of $20\times60~\mu$m$^2$ ($40\times120$~camera pixels). In our analysis, each image is split into $N$ slices along the longer side of the cloud and $\bar{\theta}_F$ is obtained for each slice, yielding $N$ magnetometry traces, as shown in Fig.~\ref{fig:detection_local}~(a). Equation~\ref{eq:mirek_trace} is fitted to each trace, providing $N$ values for the fit precision $\delta B_{0}$. Fig.~\ref{fig:detection_local}~(b) shows the mean and the standard deviation of the precision as a function of the chosen number of slices. As the number of slices $N$ grows, the determination of the magnetic field value becomes less precise, according to expectations.

This result clearly demonstrates the advantages of spatially-resolved detection. 
The chosen value of $N$ corresponds to the ability to measure the shape of the magnetic field to
the $N$-th power, which is useful when mapping complex magnetic field structures such as 
at the surface of microscopic electronic chips \cite{wildermuth_Bose_2005a}. 
In particular, we can reach a gradient precision of $3~\mu$G$/\mu$m with the vectorial magnetometer.

The spatial control of the Faraday imaging light enables spatially-selective probing in our experimental implementation. 
In particular a number of atomic samples can be probed in an array of traps, without disturbing the remaining samples.
It is therefore important to determine the detrimental effect of a given local Faraday measurement on an atomic cloud in a microtrap 
as a function of the distance between them. 
This is realised as follows:
first, a single microtrap is loaded as described above. 
Then the DMD is used to produce a Faraday probe beam with a waist of about $3.8~\mu$m, at a variable position relative to the atomic cloud.
In each experimental realisation, two sequential Faraday images 
are taken in which the position of the probe beam in the first image is variable, whereas the probe beam always hits the atomic cloud in the second image. The measurement principle is illustrated in the top panel of Fig.~\ref{fig:probe_local}~(a), where the red circle marks the position of the probe beam in the first shot, with respect to the cloud situated at a distance $s$. 
The separation $s$ is scanned and the mean rotation angle as measured from the second 
Faraday image is shown.
The effect of the first probe is clear from the reduced Faraday signal in the centre of the graph. 
The width, extracted from a Gaussian fit is $3.8(3)~\mu$m, in good 
agreement with the probe beam size. This illustrates that the effect of the probe beam is minimal when it does not hit the cloud.
The extinction ratio of the DMD
is thus good enough to completely eliminate
the need of screening of any residual light
which might originate from the Faraday light source.

Finally the potential of a spatially-selective probing scheme is further illustrated by appling the Larmor magnetometer to a triple tweezer system as shown in Fig.~\ref{fig:probe_local}~(b). The atomic clouds are separated by $13~\mu$m. In this realisation $14$ Faraday images were taken in $570~\mu$s and the traces fitted with a damped sinusoid. 
% The experiment is conducted with well calibrated fields and far from strong magnetic sources. 
The result allows us to put an upper bound on a magnetic field gradient at $\frac{dB}{dy}~=1.5~\mu$G$/\mu$m. 
This is only about one order of magnitude less precise than the best cold atoms gradiometers to date~\cite{koschorreck_High_2011,muessel_Scalable_2014a}, but in contrast to those realisations
ours can be applied in-sequence with other measurements on the same atomic cloud.

% Here the experiment was only done with $14$ Faraday pics, as we did it with a big patch. An obvious criticism is why didn't we orient the side by side on the chip and then apply 3 small patches as the probe beam. The reason is just that the loading of the dimples was a bit more homogeneous when oriented this way (due to the CDT shape) and that we hadn't tried the "locality" feature yet when we tried this out.

%%%% Outlook figure, spatial probing with quantum dynamics measurements
% tweezer scripts in: Z:\experiment\DATA_ALICE - clean\2018\02\2018-02-15\05_tweezer_array_radius_v2\tweezer_array_radius_v2\For Robert poster
% Lattice image is: oneR7_Vlatt_Piscan_v4_180615_180615_countsexp_4-2.pdf
\begin{figure*}[htbp]
	\centering
	\includegraphics[width=1\textwidth]{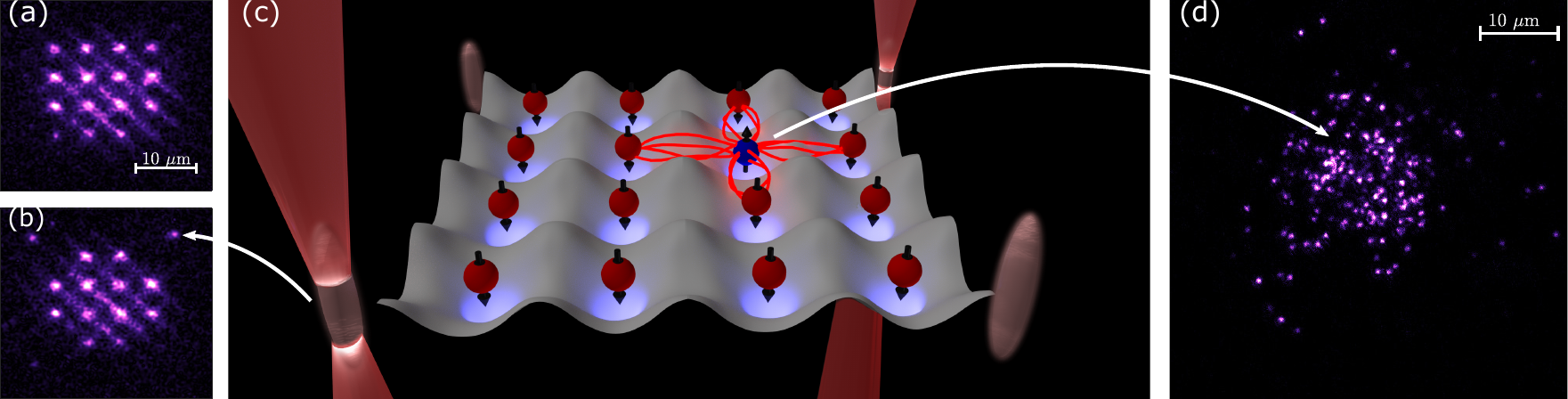}
	\caption{Proposal for enhanced quantum simulations using the spatially-selective probing scheme. (a) Fluorescence image of a 4 by 4 tweezer array of atomic clouds. The trapping potential is generated by a DMD using 940~nm laser light and projected through our newly established quantum gas microscope.  (b) Subsequently, using the dynamic capabilities of the DMD, the four corner traps are moved outwards. (c) The corner traps now can serve for instance as selective magnetic field probing regions without affecting magnetic sensitive quantum spin dynamics taking place in the centre region. (d) As an example of the interaction region, we show a fluorescence image of single atoms sparsely populating an optical lattice demonstrating the functionality of our quantum gas microscope. }\label{fig:outlook_figure}
\end{figure*}
%%%%

%%%%%%%%%%%%%%%%%%%%%%%%%%%
%%% Conclusions
%%%%%%%%%%%%%%%%%%%%%%%%%%%

\section{Outlook}
We have presented two types of magnetometers that both rely on the Faraday imaging technique. One is vectorial in nature where we reach single-shot precisions of $\delta B_{0} = 50~\mu$G and $\delta B_\perp = 100~\mu$G of the magnetic fields parallel and transverse to the direction of the Faraday probe. The achieved precision is two orders of magnitude better than a previous realisation \cite{gajdacz_Nondestructive_2013}. The second magnetometer is based on Larmor precession, making it both accurate and precise, reaching a single-shot precision of $\delta B_{y} = 15 \ \mu$G. 

The dispersive nature of the measurement technique applied allows it to be conducted 
in sequence with other experiments on cold atoms. This concept was recently demonstrated in~\cite{bason_Measurementenhanced_2018} 
and in the context of magnetometry in~\cite{krinner_situ_2018}. Our setup is equipped to create arrays of tweezers that can be individually probed without disturbing the atoms in neighbouring traps. In this way one can perform gradient or tensorial magnetometry and with improvements of the apparatus similar to~\cite{chisholm2018}, the trapping potential could also be made three-dimensional, enabling full $3$D magnetic field mapping. 
The spatially-selective aspect of the technique also allows one to perform magnetic field sensitive measurements in a portion of the system while monitoring the magnetic field in another part. In our case an atomic system could be split into arbitrarily many spatially distributed \emph{sensor regions} where the magnetic field is sensed, and an \emph{interaction region} where magnetic field sensitive dynamics take place.

% investigate measurement induced dynamics,

% In the latter paper the magnetic fields were measured via microwave spectroscopy and subsequently a 
% Rabi oscillation measurement was performed on a magnetic field sensitive hyperfine transition. 
% Fluctuations of stray magnetic fields make such measurements very difficult in an unshielded environment. 
% By correcting the population measured with the
% magnetic field value measured, the quality of the Rabi oscillation measurement improved significantly.
% old multimode was here In contrast to previous multimode probing schemes that were performed on a bulk atomic cloud~\cite{pu_Experimental_2017}, we combine the recent advances of trapping and manipulation of individual atoms in tweezer arrays and quantum gas microscopes with spatially selective dispersive probing. This means that previous limitations to quantum memory lifetimes due to atomic motion in the bulk cloud can potentially be circumvented in this approach. In addition, all the individual probing beams are collected on a camera with sufficient resolution to detect each cloud individually. This opens the possibility for simultaneous probing of several clouds, which is not possible when all signals are combined on a single detector~\cite{pu_Experimental_2017}.

% 
%As an example of how the spatially-selective measurement technique can be used is sketched in Fig.~\ref{fig:outlook_figure}. It shows a way to enhance a quantum simulation that would otherwise be limited by a fluctuating external variable, such as the magnetic field.

The experimental setup described in this paper has recently been extended to include a quantum gas microscope. It features a $0.70$~NA microscope objective capable of imaging atoms pinned in a deep 3D cubic optical lattice 
{at the resolution of a single lattice site.
Figure~\ref{fig:outlook_figure}~(d) is an example of such an image. Each bright spot corresponds to the fluorescent photons from an individual atom trapped in the lattice.}
Additionally, atoms can be trapped in flexible DMD generated conservative trapping potentials produced with $940$~nm light allowing for dynamic control of the energy landscape. The potential of this feature is shown in Fig.~\ref{fig:outlook_figure}~(a) and (b). In this experiment 16 atomic clouds were trapped in a 4 by 4 array of optical tweezers, each with a waist of about $1.0~\mu$m~(a). The corners of the array are then moved diagonally outwards by $5~\mu$m (b). As illustrated in Fig.~\ref{fig:outlook_figure}~(c), the four corners can now form four sensing regions for magnetic field measurements. They are disconnected from the central region which remains unaffected and could be used for the quantum simulation of complex many-body physics of atoms in optical lattices.

In contrast to previous multimode probing schemes that were performed on a bulk atomic cloud~\cite{pu_Experimental_2017}, limitations on quantum memory lifetimes due to atomic motion in the bulk cloud can be circumvented by such a combination of trapping and manipulation of individual atoms in tweezer arrays together with spatially-selective dispersive probing. In addition, all the individual probing beams are collected on a camera with sufficient resolution to detect each cloud individually. This opens the possibility for simultaneous probing of several clouds, which is not possible when all signals are combined on a single detector.

\begin{acknowledgments}
  The authors thank Carrie Weidner for carefully reading the manuscript, and acknowledge financial support
  from the European Research Council, the Lundbeck Foundation, the Carlsberg Foundation and the Villum Foundation.
\end{acknowledgments}

\bibliography{eliasson,SpatiallySelectiveMagnetometry}

\begin{thebibliography}{38}%
\makeatletter
\providecommand \@ifxundefined [1]{%
 \@ifx{#1\undefined}
}%
\providecommand \@ifnum [1]{%
 \ifnum #1\expandafter \@firstoftwo
 \else \expandafter \@secondoftwo
 \fi
}%
\providecommand \@ifx [1]{%
 \ifx #1\expandafter \@firstoftwo
 \else \expandafter \@secondoftwo
 \fi
}%
\providecommand \natexlab [1]{#1}%
\providecommand \enquote  [1]{``#1''}%
\providecommand \bibnamefont  [1]{#1}%
\providecommand \bibfnamefont [1]{#1}%
\providecommand \citenamefont [1]{#1}%
\providecommand \href@noop [0]{\@secondoftwo}%
\providecommand \href [0]{\begingroup \@sanitize@url \@href}%
\providecommand \@href[1]{\@@startlink{#1}\@@href}%
\providecommand \@@href[1]{\endgroup#1\@@endlink}%
\providecommand \@sanitize@url [0]{\catcode `\\12\catcode `\$12\catcode
  `\&12\catcode `\#12\catcode `\^12\catcode `\_12\catcode `\%12\relax}%
\providecommand \@@startlink[1]{}%
\providecommand \@@endlink[0]{}%
\providecommand \url  [0]{\begingroup\@sanitize@url \@url }%
\providecommand \@url [1]{\endgroup\@href {#1}{\urlprefix }}%
\providecommand \urlprefix  [0]{URL }%
\providecommand \Eprint [0]{\href }%
\providecommand \doibase [0]{http://dx.doi.org/}%
\providecommand \selectlanguage [0]{\@gobble}%
\providecommand \bibinfo  [0]{\@secondoftwo}%
\providecommand \bibfield  [0]{\@secondoftwo}%
\providecommand \translation [1]{[#1]}%
\providecommand \BibitemOpen [0]{}%
\providecommand \bibitemStop [0]{}%
\providecommand \bibitemNoStop [0]{.\EOS\space}%
\providecommand \EOS [0]{\spacefactor3000\relax}%
\providecommand \BibitemShut  [1]{\csname bibitem#1\endcsname}%
\let\auto@bib@innerbib\@empty
%</preamble>
\bibitem [{\citenamefont {Budker}\ and\ \citenamefont {{Derek F. Jackson
  Kimball}}(2013)}]{budker_Optical_2013}%
  \BibitemOpen
  \bibfield  {author} {\bibinfo {author} {\bibfnamefont {D.}~\bibnamefont
  {Budker}}\ and\ \bibinfo {author} {\bibnamefont {{Derek F. Jackson
  Kimball}}},\ }\href@noop {} {\emph {\bibinfo {title} {Optical
  {{Magnetometry}}}}}\ (\bibinfo  {publisher} {{Cambridge University Press}},\
  \bibinfo {year} {2013})\BibitemShut {NoStop}%
\bibitem [{\citenamefont {Safronova}\ \emph {et~al.}(2018)\citenamefont
  {Safronova}, \citenamefont {Budker}, \citenamefont {DeMille}, \citenamefont
  {Kimball}, \citenamefont {Derevianko},\ and\ \citenamefont
  {Clark}}]{safronova_Search_2018}%
  \BibitemOpen
  \bibfield  {author} {\bibinfo {author} {\bibfnamefont {M.~S.}\ \bibnamefont
  {Safronova}}, \bibinfo {author} {\bibfnamefont {D.}~\bibnamefont {Budker}},
  \bibinfo {author} {\bibfnamefont {D.}~\bibnamefont {DeMille}}, \bibinfo
  {author} {\bibfnamefont {D.~F.~J.}\ \bibnamefont {Kimball}}, \bibinfo
  {author} {\bibfnamefont {A.}~\bibnamefont {Derevianko}}, \ and\ \bibinfo
  {author} {\bibfnamefont {C.~W.}\ \bibnamefont {Clark}},\ }\href {\doibase
  10.1103/RevModPhys.90.025008} {\bibfield  {journal} {\bibinfo  {journal}
  {Reviews of Modern Physics}\ }\textbf {\bibinfo {volume} {90}} (\bibinfo
  {year} {2018}),\ 10.1103/RevModPhys.90.025008}\BibitemShut {NoStop}%
\bibitem [{\citenamefont {Muessel}\ \emph
  {et~al.}(2014{\natexlab{a}})\citenamefont {Muessel}, \citenamefont {Strobel},
  \citenamefont {Linnemann}, \citenamefont {Hume},\ and\ \citenamefont
  {Oberthaler}}]{muessel_Scalable_2014}%
  \BibitemOpen
  \bibfield  {author} {\bibinfo {author} {\bibfnamefont {W.}~\bibnamefont
  {Muessel}}, \bibinfo {author} {\bibfnamefont {H.}~\bibnamefont {Strobel}},
  \bibinfo {author} {\bibfnamefont {D.}~\bibnamefont {Linnemann}}, \bibinfo
  {author} {\bibfnamefont {D.~B.}\ \bibnamefont {Hume}}, \ and\ \bibinfo
  {author} {\bibfnamefont {M.~K.}\ \bibnamefont {Oberthaler}},\ }\href
  {\doibase 10.1103/PhysRevLett.113.103004} {\bibfield  {journal} {\bibinfo
  {journal} {Phys. Rev. Lett.}\ }\textbf {\bibinfo {volume} {113}},\ \bibinfo
  {pages} {103004} (\bibinfo {year} {2014}{\natexlab{a}})}\BibitemShut
  {NoStop}%
\bibitem [{\citenamefont {Kominis}\ \emph {et~al.}(2003)\citenamefont
  {Kominis}, \citenamefont {Kornack}, \citenamefont {Allred},\ and\
  \citenamefont {Romalis}}]{kominis_subfemtotesla_2003}%
  \BibitemOpen
  \bibfield  {author} {\bibinfo {author} {\bibfnamefont {I.~K.}\ \bibnamefont
  {Kominis}}, \bibinfo {author} {\bibfnamefont {T.~W.}\ \bibnamefont
  {Kornack}}, \bibinfo {author} {\bibfnamefont {J.~C.}\ \bibnamefont {Allred}},
  \ and\ \bibinfo {author} {\bibfnamefont {M.~V.}\ \bibnamefont {Romalis}},\
  }\href {\doibase 10.1038/nature01484} {\bibfield  {journal} {\bibinfo
  {journal} {Nature}\ }\textbf {\bibinfo {volume} {422}},\ \bibinfo {pages}
  {596} (\bibinfo {year} {2003})}\BibitemShut {NoStop}%
\bibitem [{\citenamefont {Dang}\ \emph {et~al.}(2010)\citenamefont {Dang},
  \citenamefont {Maloof},\ and\ \citenamefont {Romalis}}]{dang_Ultrahigh_2010}%
  \BibitemOpen
  \bibfield  {author} {\bibinfo {author} {\bibfnamefont {H.~B.}\ \bibnamefont
  {Dang}}, \bibinfo {author} {\bibfnamefont {A.~C.}\ \bibnamefont {Maloof}}, \
  and\ \bibinfo {author} {\bibfnamefont {M.~V.}\ \bibnamefont {Romalis}},\
  }\href {\doibase 10.1063/1.3491215} {\bibfield  {journal} {\bibinfo
  {journal} {Applied Physics Letters}\ }\textbf {\bibinfo {volume} {97}},\
  \bibinfo {pages} {151110} (\bibinfo {year} {2010})}\BibitemShut {NoStop}%
\bibitem [{\citenamefont {Vengalattore}\ \emph {et~al.}(2007)\citenamefont
  {Vengalattore}, \citenamefont {Higbie}, \citenamefont {Leslie}, \citenamefont
  {Guzman}, \citenamefont {Sadler},\ and\ \citenamefont
  {{Stamper-Kurn}}}]{vengalattore_HighResolution_2007}%
  \BibitemOpen
  \bibfield  {author} {\bibinfo {author} {\bibfnamefont {M.}~\bibnamefont
  {Vengalattore}}, \bibinfo {author} {\bibfnamefont {J.~M.}\ \bibnamefont
  {Higbie}}, \bibinfo {author} {\bibfnamefont {S.~R.}\ \bibnamefont {Leslie}},
  \bibinfo {author} {\bibfnamefont {J.}~\bibnamefont {Guzman}}, \bibinfo
  {author} {\bibfnamefont {L.~E.}\ \bibnamefont {Sadler}}, \ and\ \bibinfo
  {author} {\bibfnamefont {D.~M.}\ \bibnamefont {{Stamper-Kurn}}},\ }\href
  {\doibase 10.1103/PhysRevLett.98.200801} {\bibfield  {journal} {\bibinfo
  {journal} {Physical Review Letters}\ }\textbf {\bibinfo {volume} {98}}
  (\bibinfo {year} {2007}),\ 10.1103/PhysRevLett.98.200801}\BibitemShut
  {NoStop}%
\bibitem [{\citenamefont {Sewell}\ \emph {et~al.}(2012)\citenamefont {Sewell},
  \citenamefont {Koschorreck}, \citenamefont {Napolitano}, \citenamefont
  {Dubost}, \citenamefont {Behbood},\ and\ \citenamefont
  {Mitchell}}]{sewell_Magnetic_2012}%
  \BibitemOpen
  \bibfield  {author} {\bibinfo {author} {\bibfnamefont {R.~J.}\ \bibnamefont
  {Sewell}}, \bibinfo {author} {\bibfnamefont {M.}~\bibnamefont {Koschorreck}},
  \bibinfo {author} {\bibfnamefont {M.}~\bibnamefont {Napolitano}}, \bibinfo
  {author} {\bibfnamefont {B.}~\bibnamefont {Dubost}}, \bibinfo {author}
  {\bibfnamefont {N.}~\bibnamefont {Behbood}}, \ and\ \bibinfo {author}
  {\bibfnamefont {M.~W.}\ \bibnamefont {Mitchell}},\ }\href {\doibase
  10.1103/PhysRevLett.109.253605} {\bibfield  {journal} {\bibinfo  {journal}
  {Physical Review Letters}\ }\textbf {\bibinfo {volume} {109}} (\bibinfo
  {year} {2012}),\ 10.1103/PhysRevLett.109.253605}\BibitemShut {NoStop}%
\bibitem [{\citenamefont {Higbie}\ \emph {et~al.}(2005)\citenamefont {Higbie},
  \citenamefont {Sadler}, \citenamefont {Inouye}, \citenamefont {Chikkatur},
  \citenamefont {Leslie}, \citenamefont {Moore}, \citenamefont {Savalli},\ and\
  \citenamefont {{Stamper-Kurn}}}]{higbie_Direct_2005}%
  \BibitemOpen
  \bibfield  {author} {\bibinfo {author} {\bibfnamefont {J.~M.}\ \bibnamefont
  {Higbie}}, \bibinfo {author} {\bibfnamefont {L.~E.}\ \bibnamefont {Sadler}},
  \bibinfo {author} {\bibfnamefont {S.}~\bibnamefont {Inouye}}, \bibinfo
  {author} {\bibfnamefont {A.~P.}\ \bibnamefont {Chikkatur}}, \bibinfo {author}
  {\bibfnamefont {S.~R.}\ \bibnamefont {Leslie}}, \bibinfo {author}
  {\bibfnamefont {K.~L.}\ \bibnamefont {Moore}}, \bibinfo {author}
  {\bibfnamefont {V.}~\bibnamefont {Savalli}}, \ and\ \bibinfo {author}
  {\bibfnamefont {D.~M.}\ \bibnamefont {{Stamper-Kurn}}},\ }\href {\doibase
  10.1103/PhysRevLett.95.050401} {\bibfield  {journal} {\bibinfo  {journal}
  {Physical Review Letters}\ }\textbf {\bibinfo {volume} {95}} (\bibinfo {year}
  {2005}),\ 10.1103/PhysRevLett.95.050401}\BibitemShut {NoStop}%
\bibitem [{\citenamefont {Koschorreck}\ \emph {et~al.}(2011)\citenamefont
  {Koschorreck}, \citenamefont {Napolitano}, \citenamefont {Dubost},\ and\
  \citenamefont {Mitchell}}]{koschorreck_High_2011}%
  \BibitemOpen
  \bibfield  {author} {\bibinfo {author} {\bibfnamefont {M.}~\bibnamefont
  {Koschorreck}}, \bibinfo {author} {\bibfnamefont {M.}~\bibnamefont
  {Napolitano}}, \bibinfo {author} {\bibfnamefont {B.}~\bibnamefont {Dubost}},
  \ and\ \bibinfo {author} {\bibfnamefont {M.~W.}\ \bibnamefont {Mitchell}},\
  }\href {\doibase 10.1063/1.3555459} {\bibfield  {journal} {\bibinfo
  {journal} {Applied Physics Letters}\ }\textbf {\bibinfo {volume} {98}},\
  \bibinfo {pages} {074101} (\bibinfo {year} {2011})}\BibitemShut {NoStop}%
\bibitem [{\citenamefont {Muessel}\ \emph
  {et~al.}(2014{\natexlab{b}})\citenamefont {Muessel}, \citenamefont {Strobel},
  \citenamefont {Linnemann}, \citenamefont {Hume},\ and\ \citenamefont
  {Oberthaler}}]{muessel_Scalable_2014a}%
  \BibitemOpen
  \bibfield  {author} {\bibinfo {author} {\bibfnamefont {W.}~\bibnamefont
  {Muessel}}, \bibinfo {author} {\bibfnamefont {H.}~\bibnamefont {Strobel}},
  \bibinfo {author} {\bibfnamefont {D.}~\bibnamefont {Linnemann}}, \bibinfo
  {author} {\bibfnamefont {D.~B.}\ \bibnamefont {Hume}}, \ and\ \bibinfo
  {author} {\bibfnamefont {M.~K.}\ \bibnamefont {Oberthaler}},\ }\href
  {\doibase 10.1103/PhysRevLett.113.103004} {\bibfield  {journal} {\bibinfo
  {journal} {Phys. Rev. Lett.}\ }\textbf {\bibinfo {volume} {113}},\ \bibinfo
  {pages} {103004} (\bibinfo {year} {2014}{\natexlab{b}})}\BibitemShut
  {NoStop}%
\bibitem [{\citenamefont {Henderson}\ \emph {et~al.}(2009)\citenamefont
  {Henderson}, \citenamefont {Ryu}, \citenamefont {MacCormick},\ and\
  \citenamefont {Boshier}}]{Henderson2009}%
  \BibitemOpen
  \bibfield  {author} {\bibinfo {author} {\bibfnamefont {K.~C.}\ \bibnamefont
  {Henderson}}, \bibinfo {author} {\bibfnamefont {C.}~\bibnamefont {Ryu}},
  \bibinfo {author} {\bibfnamefont {C.}~\bibnamefont {MacCormick}}, \ and\
  \bibinfo {author} {\bibfnamefont {M.~G.}\ \bibnamefont {Boshier}},\ }\href
  {\doibase 10.1088/1367-2630/11/4/043030} {\bibfield  {journal} {\bibinfo
  {journal} {New J. Phys.}\ }\textbf {\bibinfo {volume} {11}},\ \bibinfo
  {pages} {043030} (\bibinfo {year} {2009})}\BibitemShut {NoStop}%
\bibitem [{\citenamefont {Endres}\ \emph {et~al.}(2016)\citenamefont {Endres},
  \citenamefont {Bernien}, \citenamefont {Keesling}, \citenamefont {Levine},
  \citenamefont {Anschuetz}, \citenamefont {Krajenbrink}, \citenamefont
  {Senko}, \citenamefont {Vuletic}, \citenamefont {Greiner},\ and\
  \citenamefont {Lukin}}]{endres_Atombyatom_2016}%
  \BibitemOpen
  \bibfield  {author} {\bibinfo {author} {\bibfnamefont {M.}~\bibnamefont
  {Endres}}, \bibinfo {author} {\bibfnamefont {H.}~\bibnamefont {Bernien}},
  \bibinfo {author} {\bibfnamefont {A.}~\bibnamefont {Keesling}}, \bibinfo
  {author} {\bibfnamefont {H.}~\bibnamefont {Levine}}, \bibinfo {author}
  {\bibfnamefont {E.~R.}\ \bibnamefont {Anschuetz}}, \bibinfo {author}
  {\bibfnamefont {A.}~\bibnamefont {Krajenbrink}}, \bibinfo {author}
  {\bibfnamefont {C.}~\bibnamefont {Senko}}, \bibinfo {author} {\bibfnamefont
  {V.}~\bibnamefont {Vuletic}}, \bibinfo {author} {\bibfnamefont
  {M.}~\bibnamefont {Greiner}}, \ and\ \bibinfo {author} {\bibfnamefont
  {M.~D.}\ \bibnamefont {Lukin}},\ }\href {\doibase 10.1126/science.aah3752}
  {\bibfield  {journal} {\bibinfo  {journal} {Science}\ }\textbf {\bibinfo
  {volume} {354}},\ \bibinfo {pages} {1024} (\bibinfo {year}
  {2016})}\BibitemShut {NoStop}%
\bibitem [{\citenamefont {Kaufman}\ \emph {et~al.}(2015)\citenamefont
  {Kaufman}, \citenamefont {Lester}, \citenamefont {{Foss-Feig}}, \citenamefont
  {Wall}, \citenamefont {Rey},\ and\ \citenamefont
  {Regal}}]{kaufman_Entangling_2015}%
  \BibitemOpen
  \bibfield  {author} {\bibinfo {author} {\bibfnamefont {A.~M.}\ \bibnamefont
  {Kaufman}}, \bibinfo {author} {\bibfnamefont {B.~J.}\ \bibnamefont {Lester}},
  \bibinfo {author} {\bibfnamefont {M.}~\bibnamefont {{Foss-Feig}}}, \bibinfo
  {author} {\bibfnamefont {M.~L.}\ \bibnamefont {Wall}}, \bibinfo {author}
  {\bibfnamefont {A.~M.}\ \bibnamefont {Rey}}, \ and\ \bibinfo {author}
  {\bibfnamefont {C.~A.}\ \bibnamefont {Regal}},\ }\href {\doibase
  10.1038/nature16073} {\bibfield  {journal} {\bibinfo  {journal} {Nature}\
  }\textbf {\bibinfo {volume} {527}},\ \bibinfo {pages} {208} (\bibinfo {year}
  {2015})}\BibitemShut {NoStop}%
\bibitem [{\citenamefont {Barredo}\ \emph {et~al.}(2016)\citenamefont
  {Barredo}, \citenamefont {{de L\'es\'eleuc}}, \citenamefont {Lienhard},
  \citenamefont {Lahaye},\ and\ \citenamefont
  {Browaeys}}]{barredo_atombyatom_2016}%
  \BibitemOpen
  \bibfield  {author} {\bibinfo {author} {\bibfnamefont {D.}~\bibnamefont
  {Barredo}}, \bibinfo {author} {\bibfnamefont {S.}~\bibnamefont {{de
  L\'es\'eleuc}}}, \bibinfo {author} {\bibfnamefont {V.}~\bibnamefont
  {Lienhard}}, \bibinfo {author} {\bibfnamefont {T.}~\bibnamefont {Lahaye}}, \
  and\ \bibinfo {author} {\bibfnamefont {A.}~\bibnamefont {Browaeys}},\ }\href
  {\doibase 10.1126/science.aah3778} {\bibfield  {journal} {\bibinfo  {journal}
  {Science}\ }\textbf {\bibinfo {volume} {354}},\ \bibinfo {pages} {1021}
  (\bibinfo {year} {2016})}\BibitemShut {NoStop}%
\bibitem [{\citenamefont {Pu}\ \emph {et~al.}(2017)\citenamefont {Pu},
  \citenamefont {Jiang}, \citenamefont {Chang}, \citenamefont {Yang},
  \citenamefont {Li},\ and\ \citenamefont {Duan}}]{pu_Experimental_2017}%
  \BibitemOpen
  \bibfield  {author} {\bibinfo {author} {\bibfnamefont {Y.-F.}\ \bibnamefont
  {Pu}}, \bibinfo {author} {\bibfnamefont {N.}~\bibnamefont {Jiang}}, \bibinfo
  {author} {\bibfnamefont {W.}~\bibnamefont {Chang}}, \bibinfo {author}
  {\bibfnamefont {H.-X.}\ \bibnamefont {Yang}}, \bibinfo {author}
  {\bibfnamefont {C.}~\bibnamefont {Li}}, \ and\ \bibinfo {author}
  {\bibfnamefont {L.-M.}\ \bibnamefont {Duan}},\ }\href {\doibase
  10.1038/ncomms15359} {\bibfield  {journal} {\bibinfo  {journal} {Nature
  Communications}\ }\textbf {\bibinfo {volume} {8}},\ \bibinfo {pages} {15359}
  (\bibinfo {year} {2017})}\BibitemShut {NoStop}%
\bibitem [{\citenamefont {Bergamini}\ \emph {et~al.}(2004)\citenamefont
  {Bergamini}, \citenamefont {Darqui\'e}, \citenamefont {Jones}, \citenamefont
  {Jacubowiez}, \citenamefont {Browaeys},\ and\ \citenamefont
  {Grangier}}]{bergamini_Holographic_2004}%
  \BibitemOpen
  \bibfield  {author} {\bibinfo {author} {\bibfnamefont {S.}~\bibnamefont
  {Bergamini}}, \bibinfo {author} {\bibfnamefont {B.}~\bibnamefont
  {Darqui\'e}}, \bibinfo {author} {\bibfnamefont {M.}~\bibnamefont {Jones}},
  \bibinfo {author} {\bibfnamefont {L.}~\bibnamefont {Jacubowiez}}, \bibinfo
  {author} {\bibfnamefont {A.}~\bibnamefont {Browaeys}}, \ and\ \bibinfo
  {author} {\bibfnamefont {P.}~\bibnamefont {Grangier}},\ }\href {\doibase
  10.1364/JOSAB.21.001889} {\bibfield  {journal} {\bibinfo  {journal} {JOSA B}\
  }\textbf {\bibinfo {volume} {21}},\ \bibinfo {pages} {1889} (\bibinfo {year}
  {2004})}\BibitemShut {NoStop}%
\bibitem [{\citenamefont {Brandt}\ \emph {et~al.}(2011)\citenamefont {Brandt},
  \citenamefont {Muldoon}, \citenamefont {Thiele}, \citenamefont {Dong},
  \citenamefont {Brainis},\ and\ \citenamefont {Kuhn}}]{brandt_Spatial_2011}%
  \BibitemOpen
  \bibfield  {author} {\bibinfo {author} {\bibfnamefont {L.}~\bibnamefont
  {Brandt}}, \bibinfo {author} {\bibfnamefont {C.}~\bibnamefont {Muldoon}},
  \bibinfo {author} {\bibfnamefont {T.}~\bibnamefont {Thiele}}, \bibinfo
  {author} {\bibfnamefont {J.}~\bibnamefont {Dong}}, \bibinfo {author}
  {\bibfnamefont {E.}~\bibnamefont {Brainis}}, \ and\ \bibinfo {author}
  {\bibfnamefont {A.}~\bibnamefont {Kuhn}},\ }\href {\doibase
  10.1007/s00340-010-4323-0} {\bibfield  {journal} {\bibinfo  {journal}
  {Applied Physics B}\ }\textbf {\bibinfo {volume} {102}},\ \bibinfo {pages}
  {443} (\bibinfo {year} {2011})}\BibitemShut {NoStop}%
\bibitem [{\citenamefont {Muldoon}(2012)}]{muldoon_Control_2012}%
  \BibitemOpen
  \bibfield  {author} {\bibinfo {author} {\bibfnamefont {C.}~\bibnamefont
  {Muldoon}},\ }\emph {\bibinfo {title} {Control and {{Manipulation}} of {{Cold
  Atoms Trapped}} in {{Optical Tweezers}}}},\ \href@noop {} {\bibinfo {type}
  {{{PhD Thesis}}}},\ \bibinfo  {school} {Oxford University}, \bibinfo
  {address} {Oxford} (\bibinfo {year} {2012})\BibitemShut {NoStop}%
\bibitem [{\citenamefont {Mazurenko}\ \emph {et~al.}(2017)\citenamefont
  {Mazurenko}, \citenamefont {Chiu}, \citenamefont {Ji}, \citenamefont
  {Parsons}, \citenamefont {{Kan\'asz-Nagy}}, \citenamefont {Schmidt},
  \citenamefont {Grusdt}, \citenamefont {Demler}, \citenamefont {Greif},\ and\
  \citenamefont {Greiner}}]{mazurenko_coldatom_2017}%
  \BibitemOpen
  \bibfield  {author} {\bibinfo {author} {\bibfnamefont {A.}~\bibnamefont
  {Mazurenko}}, \bibinfo {author} {\bibfnamefont {C.~S.}\ \bibnamefont {Chiu}},
  \bibinfo {author} {\bibfnamefont {G.}~\bibnamefont {Ji}}, \bibinfo {author}
  {\bibfnamefont {M.~F.}\ \bibnamefont {Parsons}}, \bibinfo {author}
  {\bibfnamefont {M.}~\bibnamefont {{Kan\'asz-Nagy}}}, \bibinfo {author}
  {\bibfnamefont {R.}~\bibnamefont {Schmidt}}, \bibinfo {author} {\bibfnamefont
  {F.}~\bibnamefont {Grusdt}}, \bibinfo {author} {\bibfnamefont
  {E.}~\bibnamefont {Demler}}, \bibinfo {author} {\bibfnamefont
  {D.}~\bibnamefont {Greif}}, \ and\ \bibinfo {author} {\bibfnamefont
  {M.}~\bibnamefont {Greiner}},\ }\href {\doibase 10.1038/nature22362}
  {\bibfield  {journal} {\bibinfo  {journal} {Nature}\ }\textbf {\bibinfo
  {volume} {545}},\ \bibinfo {pages} {462} (\bibinfo {year}
  {2017})}\BibitemShut {NoStop}%
\bibitem [{\citenamefont {Holland}\ \emph {et~al.}(2018)\citenamefont
  {Holland}, \citenamefont {Stuart}, \citenamefont {Barter},\ and\
  \citenamefont {Kuhn}}]{Holland2018}%
  \BibitemOpen
  \bibfield  {author} {\bibinfo {author} {\bibfnamefont {N.}~\bibnamefont
  {Holland}}, \bibinfo {author} {\bibfnamefont {D.}~\bibnamefont {Stuart}},
  \bibinfo {author} {\bibfnamefont {O.}~\bibnamefont {Barter}}, \ and\ \bibinfo
  {author} {\bibfnamefont {A.}~\bibnamefont {Kuhn}},\ }\href {\doibase
  10.1080/09500340.2018.1499978} {\bibfield  {journal} {\bibinfo  {journal} {J.
  Mod. Opt.}\ }\textbf {\bibinfo {volume} {65}},\ \bibinfo {pages} {2133}
  (\bibinfo {year} {2018})}\BibitemShut {NoStop}%
\bibitem [{\citenamefont {Simon}\ \emph {et~al.}(2011)\citenamefont {Simon},
  \citenamefont {Bakr}, \citenamefont {Ma}, \citenamefont {Tai}, \citenamefont
  {Preiss},\ and\ \citenamefont {Greiner}}]{simon_Quantum_2011}%
  \BibitemOpen
  \bibfield  {author} {\bibinfo {author} {\bibfnamefont {J.}~\bibnamefont
  {Simon}}, \bibinfo {author} {\bibfnamefont {W.~S.}\ \bibnamefont {Bakr}},
  \bibinfo {author} {\bibfnamefont {R.}~\bibnamefont {Ma}}, \bibinfo {author}
  {\bibfnamefont {M.~E.}\ \bibnamefont {Tai}}, \bibinfo {author} {\bibfnamefont
  {P.~M.}\ \bibnamefont {Preiss}}, \ and\ \bibinfo {author} {\bibfnamefont
  {M.}~\bibnamefont {Greiner}},\ }\href {\doibase 10.1038/nature09994}
  {\bibfield  {journal} {\bibinfo  {journal} {Nature}\ }\textbf {\bibinfo
  {volume} {472}},\ \bibinfo {pages} {307} (\bibinfo {year}
  {2011})}\BibitemShut {NoStop}%
\bibitem [{\citenamefont {Bason}\ \emph {et~al.}(2018)\citenamefont {Bason},
  \citenamefont {Heck}, \citenamefont {Napolitano}, \citenamefont {El\'iasson},
  \citenamefont {M\"uller}, \citenamefont {Thorsen}, \citenamefont {{Wen-Zhuo
  Zhang}}, \citenamefont {Arlt},\ and\ \citenamefont
  {Sherson}}]{bason_Measurementenhanced_2018}%
  \BibitemOpen
  \bibfield  {author} {\bibinfo {author} {\bibfnamefont {M.~G.}\ \bibnamefont
  {Bason}}, \bibinfo {author} {\bibfnamefont {R.}~\bibnamefont {Heck}},
  \bibinfo {author} {\bibfnamefont {M.}~\bibnamefont {Napolitano}}, \bibinfo
  {author} {\bibfnamefont {O.}~\bibnamefont {El\'iasson}}, \bibinfo {author}
  {\bibfnamefont {R.}~\bibnamefont {M\"uller}}, \bibinfo {author}
  {\bibfnamefont {A.}~\bibnamefont {Thorsen}}, \bibinfo {author} {\bibnamefont
  {{Wen-Zhuo Zhang}}}, \bibinfo {author} {\bibfnamefont {J.~J.}\ \bibnamefont
  {Arlt}}, \ and\ \bibinfo {author} {\bibfnamefont {J.~F.}\ \bibnamefont
  {Sherson}},\ }\href {\doibase 10.1088/1361-6455/aad447} {\bibfield  {journal}
  {\bibinfo  {journal} {Journal of Physics B: Atomic, Molecular and Optical
  Physics}\ }\textbf {\bibinfo {volume} {51}},\ \bibinfo {pages} {175301}
  (\bibinfo {year} {2018})}\BibitemShut {NoStop}%
\bibitem [{\citenamefont {Heck}\ \emph {et~al.}(2018)\citenamefont {Heck},
  \citenamefont {Vuculescu}, \citenamefont {S\o{}rensen}, \citenamefont
  {Zoller}, \citenamefont {Andreasen}, \citenamefont {Bason}, \citenamefont
  {Ejlertsen}, \citenamefont {El\'iasson}, \citenamefont {Haikka},
  \citenamefont {Laustsen}, \citenamefont {Nielsen}, \citenamefont {Mao},
  \citenamefont {M\"uller}, \citenamefont {Napolitano}, \citenamefont
  {Pedersen}, \citenamefont {Thorsen}, \citenamefont {Bergenholtz},
  \citenamefont {Calarco}, \citenamefont {Montangero},\ and\ \citenamefont
  {Sherson}}]{Heck2018}%
  \BibitemOpen
  \bibfield  {author} {\bibinfo {author} {\bibfnamefont {R.}~\bibnamefont
  {Heck}}, \bibinfo {author} {\bibfnamefont {O.}~\bibnamefont {Vuculescu}},
  \bibinfo {author} {\bibfnamefont {J.~J.}\ \bibnamefont {S\o{}rensen}},
  \bibinfo {author} {\bibfnamefont {J.}~\bibnamefont {Zoller}}, \bibinfo
  {author} {\bibfnamefont {M.~G.}\ \bibnamefont {Andreasen}}, \bibinfo {author}
  {\bibfnamefont {M.~G.}\ \bibnamefont {Bason}}, \bibinfo {author}
  {\bibfnamefont {P.}~\bibnamefont {Ejlertsen}}, \bibinfo {author}
  {\bibfnamefont {O.}~\bibnamefont {El\'iasson}}, \bibinfo {author}
  {\bibfnamefont {P.}~\bibnamefont {Haikka}}, \bibinfo {author} {\bibfnamefont
  {J.~S.}\ \bibnamefont {Laustsen}}, \bibinfo {author} {\bibfnamefont {L.~L.}\
  \bibnamefont {Nielsen}}, \bibinfo {author} {\bibfnamefont {A.}~\bibnamefont
  {Mao}}, \bibinfo {author} {\bibfnamefont {R.}~\bibnamefont {M\"uller}},
  \bibinfo {author} {\bibfnamefont {M.}~\bibnamefont {Napolitano}}, \bibinfo
  {author} {\bibfnamefont {M.~K.}\ \bibnamefont {Pedersen}}, \bibinfo {author}
  {\bibfnamefont {A.~R.}\ \bibnamefont {Thorsen}}, \bibinfo {author}
  {\bibfnamefont {C.}~\bibnamefont {Bergenholtz}}, \bibinfo {author}
  {\bibfnamefont {T.}~\bibnamefont {Calarco}}, \bibinfo {author} {\bibfnamefont
  {S.}~\bibnamefont {Montangero}}, \ and\ \bibinfo {author} {\bibfnamefont
  {J.~F.}\ \bibnamefont {Sherson}},\ }\href {\doibase 10.1073/pnas.1716869115}
  {\bibfield  {journal} {\bibinfo  {journal} {Proc. Natl. Acad. Sci.}\ }\textbf
  {\bibinfo {volume} {115}},\ \bibinfo {pages} {E11231} (\bibinfo {year}
  {2018})}\BibitemShut {NoStop}%
\bibitem [{\citenamefont {Gajdacz}\ \emph {et~al.}(2013)\citenamefont
  {Gajdacz}, \citenamefont {Pedersen}, \citenamefont {M\o{}rch}, \citenamefont
  {Hilliard}, \citenamefont {Arlt},\ and\ \citenamefont
  {Sherson}}]{gajdacz_Nondestructive_2013}%
  \BibitemOpen
  \bibfield  {author} {\bibinfo {author} {\bibfnamefont {M.}~\bibnamefont
  {Gajdacz}}, \bibinfo {author} {\bibfnamefont {P.~L.}\ \bibnamefont
  {Pedersen}}, \bibinfo {author} {\bibfnamefont {T.}~\bibnamefont {M\o{}rch}},
  \bibinfo {author} {\bibfnamefont {A.~J.}\ \bibnamefont {Hilliard}}, \bibinfo
  {author} {\bibfnamefont {J.}~\bibnamefont {Arlt}}, \ and\ \bibinfo {author}
  {\bibfnamefont {J.~F.}\ \bibnamefont {Sherson}},\ }\href {\doibase
  http://dx.doi.org/10.1063/1.4818913} {\bibfield  {journal} {\bibinfo
  {journal} {Rev. Sci. Instrum.}\ }\textbf {\bibinfo {volume} {84}} (\bibinfo
  {year} {2013}),\ http://dx.doi.org/10.1063/1.4818913}\BibitemShut {NoStop}%
\bibitem [{\citenamefont {Isayama}\ \emph {et~al.}(1999)\citenamefont
  {Isayama}, \citenamefont {Takahashi}, \citenamefont {Tanaka}, \citenamefont
  {Toyoda}, \citenamefont {Ishikawa},\ and\ \citenamefont
  {Yabuzaki}}]{isayama_Observation_1999}%
  \BibitemOpen
  \bibfield  {author} {\bibinfo {author} {\bibfnamefont {T.}~\bibnamefont
  {Isayama}}, \bibinfo {author} {\bibfnamefont {Y.}~\bibnamefont {Takahashi}},
  \bibinfo {author} {\bibfnamefont {N.}~\bibnamefont {Tanaka}}, \bibinfo
  {author} {\bibfnamefont {K.}~\bibnamefont {Toyoda}}, \bibinfo {author}
  {\bibfnamefont {K.}~\bibnamefont {Ishikawa}}, \ and\ \bibinfo {author}
  {\bibfnamefont {T.}~\bibnamefont {Yabuzaki}},\ }\href {\doibase
  10.1103/PhysRevA.59.4836} {\bibfield  {journal} {\bibinfo  {journal}
  {Physical Review A}\ }\textbf {\bibinfo {volume} {59}},\ \bibinfo {pages}
  {4836} (\bibinfo {year} {1999})}\BibitemShut {NoStop}%
\bibitem [{\citenamefont {Palacios}\ \emph {et~al.}(2018)\citenamefont
  {Palacios}, \citenamefont {Coop}, \citenamefont {Gomez}, \citenamefont
  {Vanderbruggen}, \citenamefont {de~Escobar}, \citenamefont {{Martijn
  Jasperse}},\ and\ \citenamefont {Mitchell}}]{palacios_Multisecond_2018}%
  \BibitemOpen
  \bibfield  {author} {\bibinfo {author} {\bibfnamefont {S.}~\bibnamefont
  {Palacios}}, \bibinfo {author} {\bibfnamefont {S.}~\bibnamefont {Coop}},
  \bibinfo {author} {\bibfnamefont {P.}~\bibnamefont {Gomez}}, \bibinfo
  {author} {\bibfnamefont {T.}~\bibnamefont {Vanderbruggen}}, \bibinfo {author}
  {\bibfnamefont {Y.~N.~M.}\ \bibnamefont {de~Escobar}}, \bibinfo {author}
  {\bibnamefont {{Martijn Jasperse}}}, \ and\ \bibinfo {author} {\bibfnamefont
  {M.~W.}\ \bibnamefont {Mitchell}},\ }\href {\doibase
  10.1088/1367-2630/aab2a0} {\bibfield  {journal} {\bibinfo  {journal} {New
  Journal of Physics}\ }\textbf {\bibinfo {volume} {20}},\ \bibinfo {pages}
  {053008} (\bibinfo {year} {2018})}\BibitemShut {NoStop}%
\bibitem [{\citenamefont {Kaminski}\ \emph {et~al.}(2012)\citenamefont
  {Kaminski}, \citenamefont {Kampel}, \citenamefont {Steenstrup}, \citenamefont
  {Griesmaier}, \citenamefont {Polzik},\ and\ \citenamefont
  {M\"uller}}]{kaminski_Insitu_2012}%
  \BibitemOpen
  \bibfield  {author} {\bibinfo {author} {\bibfnamefont {F.}~\bibnamefont
  {Kaminski}}, \bibinfo {author} {\bibfnamefont {N.~S.}\ \bibnamefont
  {Kampel}}, \bibinfo {author} {\bibfnamefont {M.~P.~H.}\ \bibnamefont
  {Steenstrup}}, \bibinfo {author} {\bibfnamefont {A.}~\bibnamefont
  {Griesmaier}}, \bibinfo {author} {\bibfnamefont {E.~S.}\ \bibnamefont
  {Polzik}}, \ and\ \bibinfo {author} {\bibfnamefont {J.~H.}\ \bibnamefont
  {M\"uller}},\ }\href {\doibase 10.1140/epjd/e2012-30038-0} {\bibfield
  {journal} {\bibinfo  {journal} {The European Physical Journal D}\ }\textbf
  {\bibinfo {volume} {66}} (\bibinfo {year} {2012}),\
  10.1140/epjd/e2012-30038-0}\BibitemShut {NoStop}%
\bibitem [{\citenamefont {Gajdacz}\ \emph {et~al.}(2016)\citenamefont
  {Gajdacz}, \citenamefont {Hilliard}, \citenamefont {Kristensen},
  \citenamefont {Pedersen}, \citenamefont {Klempt}, \citenamefont {Arlt},\ and\
  \citenamefont {Sherson}}]{gajdacz_Preparation_2016}%
  \BibitemOpen
  \bibfield  {author} {\bibinfo {author} {\bibfnamefont {M.}~\bibnamefont
  {Gajdacz}}, \bibinfo {author} {\bibfnamefont {A.~J.}\ \bibnamefont
  {Hilliard}}, \bibinfo {author} {\bibfnamefont {M.~A.}\ \bibnamefont
  {Kristensen}}, \bibinfo {author} {\bibfnamefont {P.~L.}\ \bibnamefont
  {Pedersen}}, \bibinfo {author} {\bibfnamefont {C.}~\bibnamefont {Klempt}},
  \bibinfo {author} {\bibfnamefont {J.~J.}\ \bibnamefont {Arlt}}, \ and\
  \bibinfo {author} {\bibfnamefont {J.~F.}\ \bibnamefont {Sherson}},\ }\href
  {\doibase 10.1103/PhysRevLett.117.073604} {\bibfield  {journal} {\bibinfo
  {journal} {Physical Review Letters}\ }\textbf {\bibinfo {volume} {117}}
  (\bibinfo {year} {2016}),\ 10.1103/PhysRevLett.117.073604}\BibitemShut
  {NoStop}%
\bibitem [{\citenamefont {Bradley}\ \emph {et~al.}(1997)\citenamefont
  {Bradley}, \citenamefont {Sackett},\ and\ \citenamefont
  {Hulet}}]{bradley_BoseEinstein_1997}%
  \BibitemOpen
  \bibfield  {author} {\bibinfo {author} {\bibfnamefont {C.~C.}\ \bibnamefont
  {Bradley}}, \bibinfo {author} {\bibfnamefont {C.~A.}\ \bibnamefont
  {Sackett}}, \ and\ \bibinfo {author} {\bibfnamefont {R.~G.}\ \bibnamefont
  {Hulet}},\ }\href {\doibase 10.1103/PhysRevLett.78.985} {\bibfield  {journal}
  {\bibinfo  {journal} {Physical Review Letters}\ }\textbf {\bibinfo {volume}
  {78}},\ \bibinfo {pages} {985} (\bibinfo {year} {1997})}\BibitemShut
  {NoStop}%
\bibitem [{Note1()}]{Note1}%
  \BibitemOpen
  \bibinfo {note} {{Note, that one is not necessarily limited to constant sweep
  rates. Reducing for example the sweep rate around the zero crossing of the
  Faraday signal could increase the precision of the magnetometer
  further.}}\BibitemShut {Stop}%
\bibitem [{\citenamefont {{Dupont-Roc}}\ \emph {et~al.}(1969)\citenamefont
  {{Dupont-Roc}}, \citenamefont {Haroche},\ and\ \citenamefont
  {{Cohen-Tannoudji}}}]{dupont-roc_Detection_1969}%
  \BibitemOpen
  \bibfield  {author} {\bibinfo {author} {\bibfnamefont {J.}~\bibnamefont
  {{Dupont-Roc}}}, \bibinfo {author} {\bibfnamefont {S.}~\bibnamefont
  {Haroche}}, \ and\ \bibinfo {author} {\bibfnamefont {C.}~\bibnamefont
  {{Cohen-Tannoudji}}},\ }\href {\doibase 10.1016/0375-9601(69)90480-0}
  {\bibfield  {journal} {\bibinfo  {journal} {Physics Letters A}\ }\textbf
  {\bibinfo {volume} {28}},\ \bibinfo {pages} {638} (\bibinfo {year}
  {1969})}\BibitemShut {NoStop}%
\bibitem [{\citenamefont {Smith}\ \emph {et~al.}(2011)\citenamefont {Smith},
  \citenamefont {Anderson}, \citenamefont {Chaudhury},\ and\ \citenamefont
  {Jessen}}]{smith_Threeaxis_2011}%
  \BibitemOpen
  \bibfield  {author} {\bibinfo {author} {\bibfnamefont {A.}~\bibnamefont
  {Smith}}, \bibinfo {author} {\bibfnamefont {B.~E.}\ \bibnamefont {Anderson}},
  \bibinfo {author} {\bibfnamefont {S.}~\bibnamefont {Chaudhury}}, \ and\
  \bibinfo {author} {\bibfnamefont {P.~S.}\ \bibnamefont {Jessen}},\ }\href
  {\doibase 10.1088/0953-4075/44/20/205002} {\bibfield  {journal} {\bibinfo
  {journal} {Journal of Physics B: Atomic, Molecular and Optical Physics}\
  }\textbf {\bibinfo {volume} {44}},\ \bibinfo {pages} {205002} (\bibinfo
  {year} {2011})}\BibitemShut {NoStop}%
\bibitem [{\citenamefont {Krinner}\ \emph {et~al.}(2018)\citenamefont
  {Krinner}, \citenamefont {Stewart}, \citenamefont {Pazmi\~no},\ and\
  \citenamefont {Schneble}}]{krinner_situ_2018}%
  \BibitemOpen
  \bibfield  {author} {\bibinfo {author} {\bibfnamefont {L.}~\bibnamefont
  {Krinner}}, \bibinfo {author} {\bibfnamefont {M.}~\bibnamefont {Stewart}},
  \bibinfo {author} {\bibfnamefont {A.}~\bibnamefont {Pazmi\~no}}, \ and\
  \bibinfo {author} {\bibfnamefont {D.}~\bibnamefont {Schneble}},\ }\href
  {\doibase 10.1063/1.5003646} {\bibfield  {journal} {\bibinfo  {journal}
  {Review of Scientific Instruments}\ }\textbf {\bibinfo {volume} {89}},\
  \bibinfo {pages} {013108} (\bibinfo {year} {2018})}\BibitemShut {NoStop}%
\bibitem [{Note2()}]{Note2}%
  \BibitemOpen
  \bibinfo {note} {{In the field of ultracold quantum gases, magnetic fields
  are typically given in units of Gauss, and use this convention here, as the
  paper is primarily intended for that audience. However, this choice of units
  is not the same in the field of optical magnetometry, where the central
  figure of merit, the sensitivity, is typically given in SI units (of
  T~Hz$^{-\protect \frac {1}{2}}$). For this reason we state magnetic fields in
  units of G, but sensitivities in units of T~Hz$^{-\protect \frac
  {1}{2}}$.}}\BibitemShut {Stop}%
\bibitem [{\citenamefont {Steck}(2010)}]{steck_Rubidium_2010}%
  \BibitemOpen
  \bibfield  {author} {\bibinfo {author} {\bibfnamefont {D.}~\bibnamefont
  {Steck}},\ }\href@noop {} {\emph {\bibinfo {title} {Rubidium 87 {{D Line
  Data}}}}},\ \bibinfo {edition} {2nd}\ ed.\ (\bibinfo {year}
  {2010})\BibitemShut {NoStop}%
\bibitem [{\citenamefont {Fatemi}\ and\ \citenamefont
  {Bashkansky}(2010)}]{fatemi_Spatially_2010}%
  \BibitemOpen
  \bibfield  {author} {\bibinfo {author} {\bibfnamefont {F.~K.}\ \bibnamefont
  {Fatemi}}\ and\ \bibinfo {author} {\bibfnamefont {M.}~\bibnamefont
  {Bashkansky}},\ }\href {\doibase 10.1364/OE.18.002190} {\bibfield  {journal}
  {\bibinfo  {journal} {Optics Express}\ }\textbf {\bibinfo {volume} {18}},\
  \bibinfo {pages} {2190} (\bibinfo {year} {2010})}\BibitemShut {NoStop}%
\bibitem [{\citenamefont {Wildermuth}\ \emph {et~al.}(2005)\citenamefont
  {Wildermuth}, \citenamefont {Hofferberth}, \citenamefont {Lesanovsky},
  \citenamefont {Haller}, \citenamefont {Andersson}, \citenamefont {Groth},
  \citenamefont {{Bar-Joseph}}, \citenamefont {Kr\"uger},\ and\ \citenamefont
  {Schmiedmayer}}]{wildermuth_Bose_2005a}%
  \BibitemOpen
  \bibfield  {author} {\bibinfo {author} {\bibfnamefont {S.}~\bibnamefont
  {Wildermuth}}, \bibinfo {author} {\bibfnamefont {S.}~\bibnamefont
  {Hofferberth}}, \bibinfo {author} {\bibfnamefont {I.}~\bibnamefont
  {Lesanovsky}}, \bibinfo {author} {\bibfnamefont {E.}~\bibnamefont {Haller}},
  \bibinfo {author} {\bibfnamefont {L.~M.}\ \bibnamefont {Andersson}}, \bibinfo
  {author} {\bibfnamefont {S.}~\bibnamefont {Groth}}, \bibinfo {author}
  {\bibfnamefont {I.}~\bibnamefont {{Bar-Joseph}}}, \bibinfo {author}
  {\bibfnamefont {P.}~\bibnamefont {Kr\"uger}}, \ and\ \bibinfo {author}
  {\bibfnamefont {J.}~\bibnamefont {Schmiedmayer}},\ }\href {\doibase
  10.1038/435440a} {\bibfield  {journal} {\bibinfo  {journal} {Nature}\
  }\textbf {\bibinfo {volume} {435}},\ \bibinfo {pages} {440} (\bibinfo {year}
  {2005})}\BibitemShut {NoStop}%
\bibitem [{\citenamefont {Chisholm}\ \emph {et~al.}(2018)\citenamefont
  {Chisholm}, \citenamefont {Thomas}, \citenamefont {Deb},\ and\ \citenamefont
  {Kj\ae{}rgaard}}]{chisholm2018}%
  \BibitemOpen
  \bibfield  {author} {\bibinfo {author} {\bibfnamefont {C.~S.}\ \bibnamefont
  {Chisholm}}, \bibinfo {author} {\bibfnamefont {R.}~\bibnamefont {Thomas}},
  \bibinfo {author} {\bibfnamefont {A.~B.}\ \bibnamefont {Deb}}, \ and\
  \bibinfo {author} {\bibfnamefont {N.}~\bibnamefont {Kj\ae{}rgaard}},\ }\href
  {\doibase 10.1063/1.5041481} {\bibfield  {journal} {\bibinfo  {journal} {Rev.
  Sci. Instrum.}\ }\textbf {\bibinfo {volume} {89}},\ \bibinfo {pages} {103105}
  (\bibinfo {year} {2018})}\BibitemShut {NoStop}%
\end{thebibliography}%
\end{document}